\documentclass[journal]{IEEEtran}
\usepackage{graphicx}
\usepackage{textcomp}
\usepackage{xcolor}
\usepackage{colortbl}
\definecolor{mygray}{gray}{0.98}
\definecolor{mygray1}{gray}{0.93}
\usepackage{algorithm}  
\usepackage{algpseudocode}  
\usepackage{amsmath}  
\usepackage{amsfonts}
\usepackage{amssymb}
\usepackage{latexsym}
\usepackage{caption, subcaption}
\usepackage{enumitem}
\usepackage{booktabs}
\usepackage{listings}
\usepackage{chngpage}
\usepackage{flushend}
\usepackage{makecell}
\usepackage{adjustbox}
\usepackage{bbding}
\usepackage{utfsym}
\usepackage{autobreak}
\usepackage{multirow}
\usepackage{siunitx}
\usepackage{textcomp,booktabs}
\usepackage{wasysym}

\usepackage{ragged2e}
\definecolor{seagreen}{rgb}{0.18, 0.55, 0.34}
\definecolor{royalpurple}{rgb}{0.47,0.32,0.66}
\definecolor{brown(traditional)}{rgb}{0.59, 0.29, 0.0}
\definecolor{blue}{rgb}{0.3, 0.2, 0.9}
\usepackage[colorlinks,
            linkcolor=blue,
            anchorcolor=blue,
            citecolor=blue]{hyperref}
\ifCLASSINFOpdf
\else
\fi
\begin{document}

%
\title{Cross-Modal Generative Semantic Communications for Mobile AIGC: Joint Semantic Encoding and Prompt Engineering}
%
%
%

\author{Yinqiu~Liu,
        Hongyang~Du,
        Dusit~Niyato,~\IEEEmembership{Fellow,~IEEE},
        Jiawen~Kang,
        Zehui~Xiong,
        Shiwen~Mao,~\IEEEmembership{Fellow,~IEEE},
        Ping~Zhang,~\IEEEmembership{Fellow,~IEEE},
        and Xuemin (Sherman) Shen,~\IEEEmembership{Fellow,~IEEE}
\thanks{Y. Liu, H. Du, and D. Niyato are with the School of Computer Science and Engineering, Nanyang Technological University, Singapore. (e-mail:yinqiu001@e.ntu.edu.sg, hongyang001@e.ntu.edu.sg, dniyato@ntu.edu.sg).}
\thanks{J. Kang is with the School of Automation, Guangdong University of Technology, China (e-mail: kavinkang@gdut.edu.cn).}
\thanks{Z. Xiong is with the Pillar of Information Systems Technology and Design, Singapore University of Technology and Design, Singapore (e-mail: zehui xiong@sutd.edu.sg).}
\thanks{S. Mao is with the Department of Electrical and Computer Engineering, Auburn University, Auburn, USA (e-mail: smao@ieee.org)}
\thanks{P. Zhang is with the State Key Laboratory of Networking and Switching Technology, Beijing University of Posts and Telecommunications, China (e-mail: pzhang@bupt.edu.cn).}
\thanks{X. Shen is with the Department of Electrical and Computer Engineering, University of Waterloo, Canada (email: sshen@uwaterloo.ca).}
}

\maketitle

\begin{abstract}
Employing massive Mobile AI-Generated Content (AIGC) Service Providers (MASPs) with powerful models, high-quality AIGC services can become accessible for resource-constrained end users. 
However, this advancement, referred to as mobile AIGC, also introduces a significant challenge: users should download large AIGC outputs from the MASPs, leading to substantial bandwidth consumption and potential transmission failures.
In this paper, we apply cross-modal \underline{G}enerative \underline{Sem}antic \underline{Com}munications (G-SemCom) in mobile AIGC to overcome wireless bandwidth constraints.
Specifically, we utilize a series of cross-modal attention maps to indicate the correlation between user prompts and each part of AIGC outputs.
In this way, the MASP can analyze the prompt context and filter the most semantically important content efficiently. 
Only semantic information is transmitted, with which users can recover the entire AIGC output with high quality while saving mobile bandwidth.
Since the transmitted information not only preserves the semantics but also prompts the recovery, we formulate a joint semantic encoding and prompt engineering problem to optimize the bandwidth allocation among users. 
Particularly, we present a human-perceptual metric named Joint Perpetual Similarity and Quality (JPSQ), which is fused by two learning-based measurements regarding semantic similarity and aesthetic quality, respectively.
Furthermore, we develop the Attention-aware Deep Diffusion (ADD) algorithm, which learns attention maps and leverages the diffusion process to enhance the environment exploration ability of traditional deep reinforcement learning (DRL). 
Extensive experiments demonstrate that our proposal can reduce the bandwidth consumption of mobile users by 49.4\% on average, with almost no perceptual difference in AIGC output quality. 
Moreover, the ADD algorithm shows superior performance over baseline DRL methods, with 1.74$\times$ higher overall reward. 
\end{abstract}


%
\IEEEpeerreviewmaketitle

\section{Introduction}
%
%
%
%
\IEEEPARstart{A}{s} the latest paradigm for content creation in the Metaverse era, AI-Generated Content (AIGC) \cite{xu2023unleashing,du2023beyond} has attracted great attention from both academia and industry.
Recently, we have witnessed the phenomenal success of AIGC in various fields, such as Stable Diffusion and DALL-E$\cdot$3 in text-to-image generation, ChatGPT in Q \& A, and MusicLM in music composition \cite{liu2023optimizing}.
Nonetheless, the strong power of AIGC models relies on extremely large neural networks with billions of learnable parameters.
For instance, DALL-E$\cdot$2 and GPT-3 contain 3.5 billion and 175 billion parameters, respectively \cite{liu2023optimizing}.
Moreover, considering the difficulty of generating content following complex distributions, such as images and videos, each round of generative inference costs considerable power.
Such resource-intensive features severely hinder the further application of AIGC, especially in mobile/edge scenarios with resource constraints.

To overcome resource limitations and provide ubiquitous high-quality AIGC services, researchers sought help from mobile-edge computing and presented the concept of \textit{Mobile AIGC} \cite{xu2023unleashing}.
Specifically, massive end users can offload their AIGC tasks to Mobile AIGC Service Providers (MASPs), e.g., base stations.
With abundant computing resources to operate AIGC models, the MASP can provide paid AIGC inference services according to users' task description/input, so-called prompts.
In this way, users can receive high-quality AIGC outputs while circumventing hefty computation costs on their mobile devices.
Moreover, mobile communications protect the users' privacy, avoiding numerous users transmitting sensitive prompts (e.g., containing personal information) to a remote cloud AIGC server.
Recently, a series of breakthroughs regarding optimizing AIGC models and managing mobile AIGC networks have been proposed.
For instance, Qualcomm published the world's first on-device Stable Diffusion \cite{qualcomm}.
Chen \textit{et al.} \cite{chen2023speed} performed GPU-aware optimization on large diffusion models, accomplishing fast text-to-image AIGC on mobile-edge servers and devices.
From the network perspective, Du \textit{et al.} \cite{10172151} and Wen \textit{et al.} \cite{10233667} scheduled the task allocation between users and the MASPs and designed the incentive mechanism for mobile AIGC, respectively.

Despite the achievements that have been made, the existing works ignore the bandwidth consumption of mobile devices.
We observe that mobile AIGC just reduces users' computation overhead at the expense of increasing bandwidth consumption since users should download large AIGC outputs from the MASP after each round of inference.
Hence, two challenges exist in the current paradigm.
\begin{itemize}
    \item \textbf{Modality Transfer during AIGC Inference}: AIGC inference generally involves generating high-dimension information from low-dimension prompts, e.g., generating images (hundreds of KBs) from texts (hundreds of bytes). Such modality transfers might cause failed transmission if large AIGC outputs block the downlink channel. Incomplete or damaged AIGC outputs are less useful to users and downstream applications.
    \item \textbf{Contradiction between Generation Quality and Bandwidth Consumption}: The higher the quality of AIGC outputs, typically, the larger their sizes. Accordingly, more bandwidth is required for transmission. Therefore, if encountering transmission failure, users need to adjust and/or reduce their requirements for the quality of AIGC outputs and ask the MASP for regeneration. Such a contradiction prevents users from receiving high-quality AIGC outputs. Moreover, regeneration consumes additional bandwidth.
\end{itemize}

In this paper, we introduce Semantic Communications (SemCom) \cite{Yining} in mobile AIGC to overcome the bandwidth constraints.
Instead of transmitting every bit, SemCom circumvents the channel capacity limitation by only transmitting critical semantic information, enabling users to accomplish specific applications while saving wireless bandwidth.
In SemCom-aided mobile AIGC, a MASP can extract semantic features of the AIGC outputs, thereby compressing the content to be transmitted.
Then, the users can apply a lightweight decoder to recover the source AIGC outputs with high fidelity.
Note that several studies have explored the potential of SemCom in mobile AIGC \cite{Add1, Add2}. 
However, they do not realize the systematic SemCom-aided mobile AIGC and perform intensive experiments to illustrate how much bandwidth can be saved by SemCom without affecting the AIGC output quality on the user side. 
In contrast, we present a novel SemCom framework containing three mobile AIGC-oriented designs:
i) To extract AIGC outputs' semantics and perform output recovery efficiently, we introduce a semantic extraction module in the MASP's AIGC model and equip users with generative decoders, forming the Generative SemCom (G-SemCom).
ii) Noticing the modality transfers during AIGC inferences, our semantic information takes the form of a series of cross-modal attention maps, which associate each prompt word to certain parts of the AIGC output by attention scores. Hence, we can perform fine-grained semantic analysis of the AIGC outputs, filtering the content with the most important semantic meaning for users.
iii) Traditional SemCom only optimizes the semantic similarity between the source and recovered information. However, in mobile AIGC, users require the recovered AIGC outputs to be high-quality. To this end, we present the joint optimization, which performs prompt engineering to ensure output quality when allocating wireless bandwidth for output transmission.  

The main contributions of this paper can be summarized as follows:
\begin{itemize}
    \item \textbf{G-SemCom Framework for Mobile AIGC}: \textit{To the best of our knowledge, we are the first to present the cross-modal G-SemCom framework for mobile AIGC.} Supported by G-SemCom, a MASP only needs to transmit compressed semantic information of the AIGC output. On the user side, a generative decoder is deployed for recovery. In this way, the users can acquire high-quality AIGC outputs while saving considerable computation and bandwidth resources.
    \item \textbf{Attention-Aware Semantic Extraction}: Noticing the cross-modality feature of mobile AIGC, we propose an attention-aware method to extract semantic features of the source information. Specifically, we visualize the activation of the cross-attention layers in diffusion-based AIGC models, forming a series of cross-modal attention maps. By scoring the correlation between the user prompt and each part of the generated AIGC output, efficient semantic encoding can be performed from the user's perspective, thereby ensuring the semantic correctness of the recovered AIGC output.
    \item \textbf{Joint Semantic Encoding and Prompt Engineering}: We formulate a joint optimization problem to optimize the bandwidth allocation. Particularly, since the information sent by MASP not only preserves semantic features but also serves as the prompt for guiding the recovery of AIGC outputs, we consider the joint semantic encoding and prompt engineering with the goal of simultaneously maximizing the semantic similarity and output quality while saving wireless bandwidth. To do so, we define a novel human-perceptual metric called Joint Perpetual Similarity and Quality (JPSQ) to indicate the efficiency of G-SemCom in mobile AIGC. Moreover, we develop the Attention-aware Deep Diffusion (ADD) algorithm to solve the optimization, which utilizes diffusion steps to achieve strong exploration ability.
    \item \textbf{Experimental Results}: Extensive experiments prove the validity of our proposals. Specifically, the bandwidth consumption of mobile users can be reduced by 49.4\% on average, while the perpetual output quality score \cite{NIMA} only drops by 0.0299. Moreover, the ADD algorithm significantly outperforms baseline Deep Reinforcement Learning (DRL) algorithms regarding converge speed and efficiency for bandwidth allocation.
\end{itemize}

The remainder of this paper is organized as follows.
Section II reviews the related works.
Our motivation and system model are shown in Section III.
In Section IV, we elaborate on our G-SemCom framework, especially the attention-aware semantic extraction, for mobile AIGC.
Then, Section V describes the joint optimization of semantic encoding and prompt engineering.
The experimental results and analysis are illustrated in Section VI.
Finally, Section VII concludes the paper.

\section{Related Work}
\subsection{Mobile AIGC}
Recently, mobile AIGC has attracted a lot of attention from both the model and system aspects \cite{xu2023unleashing}.
Firstly, a series of lightweight AIGC models have been presented.
For instance, Chen \textit{et al.} \cite{chen2023speed} conducted GPU-aware optimization on large diffusion models, realizing the on-device text-to-image generation in 12 seconds.
SnapFusion \cite{li2023snapfusion} utilized step distillation and further reduced the inference time to 2 seconds on mobile devices.
In addition to academia, the industry also paid great attention to improving the adaptability of AIGC models and attracting more mobile users.
On Feb. 2023, Qualcomm developed the world's first on-device Stable Diffusion \cite{qualcomm}.
Likewise, Google and Apple also presented MediaPipe \cite{Google} and Core-ML Stable Diffusion \cite{Apple}, respectively.
From the system perspective, the architecture and management of mobile AIGC are evolving rapidly.
Xu \textit{et al.} \cite{xu2023unleashing} systemically introduced the potential of mobile-edge networks for accommodating AIGC services. 
Du \textit{et al.} \cite{10172151} discussed the task scheduling of mobile-edge AIGC, improving the system capacity by assigning each AIGC task to the most appropriate MASP.
Wen \textit{et al.} \cite{10233667} designed the incentive mechanism for rewarding MASPs, ensuring the participation and economic sustainability of mobile AIGC.
By offloading AIGC tasks to MASPs, the computation resource consumption of users can be greatly decreased. 
However, users need to frequently download large AIGC outputs from MASP, which costs huge wireless bandwidth. 
To this end, we present an end-to-end SemCom framework for mobile AIGC.

\subsection{Generative Semantic Communications (G-SemCom)}
Generative models have shown great potential to be incorporated into SemCom, which we coin as G-SemCom \cite{grassucci2023generative}.
On the sender side, generative models can help extract human-interpretable semantic features.
For instance, Wang \textit{et al.} \cite{Yining} generated semantic triples (formed by \textit{object A-relationship-object B}) to summarize and compress the content of the source textual message.
For image-oriented G-SemCom, Liu \textit{et al.} \cite{liu2023semantic} evaluated various forms for representing visual semantics, e.g., skeleton and depth maps.
Compared with parameterized semantic features used by traditional SemCom, human-interpretable triples and images are easy to analyze, making human-in-the-loop SemCom optimization possible.
On the receiver side, efficient semantic decoding and information recovery can be easily realized by generative decoders due to their outstanding generation ability.
For instance, He \textit{et al.} \cite{10000735} and Grassucci \textit{et al.} \cite{grassucci2023generative} adopted generative adversarial networks and diffusion models to reconstruct images that are semantically
equivalent to the source images.
Furthermore, Du \textit{et al.} \cite{du2023enabling2} explored the multi-modal G-SemCom, realizing the bit-level accurate recovery. 
Finally, since generative models are trained on huge datasets and master massive knowledge, they can serve as powerful shared knowledge bases between senders and receivers.
For example, Jiang \textit{et al.} \cite{jiang2023large} developed a training-free knowledge base by Meta Segment Anything, realizing zero-shot semantic extraction and supporting fast information recovery.
Despite the achievement in the G-SemCom field, none of these works are targeted to mobile AIGC.
In mobile AIGC, the receivers consider not only the semantic similarity but also the ``quality" of the recovered AIGC output.
Moreover, since AIGC usually involves content creation across modalities, e.g., text-to-image generations, the cross-modal semantic mapping should be explored when designing G-SemCom.

\subsection{Prompt Engineering}
Prompt engineering refers to the process of crafting/finding the most appropriate prompt for the given downstream task, aiming to maximize the generation quality \cite{10.1145/3560815}.
According to the specific prompt form, various prompt engineering methods have been proposed.
For textual prompts, e.g., the instructions that we input to ChatGPT, the authors in \cite{haviv-etal-2021-bertese, wallace-etal-2019-universal}, and \cite{gao-etal-2021-making} presented the prompt paraphrasing, searching, and generation, respectively.
Despite adopting different strategies, the common principle is finding the best textual template for reformulating raw prompts, facilitating PFMs to associate downstream tasks with the learned knowledge.
The parameterized prompts, e.g., the input preprocessing layers appended to PFMs, can be optimized by prompt-oriented tuning \cite{tanwisuth2023pouf}. 
Moreover, Guo \textit{et al.} \cite{10210127} and Zhao \textit{et al.} \cite{10095356} incorporated prompt engineering in federated learning, which allows multiple users to contribute data and physical resources and optimize the prompts collaboratively.
Apart from determining the generation quality, in mobile AIGC, prompt engineering also directly affects the network-level performance.
Liu \textit{et al.} \cite{liu2023optimizing} illustrated that if users keep using low-quality prompts, frequent re-generation will cause considerable service fees, service latency, and bandwidth consumption.
Noticing the importance of prompt engineering, we jointly consider the semantic encoding and prompt engineering on the sender side for the first time, thereby optimizing the input fed to the receiver's generative decoder while saving bandwidth.
\begin{figure}[tbp]
\centerline{\includegraphics[width=0.98\columnwidth]{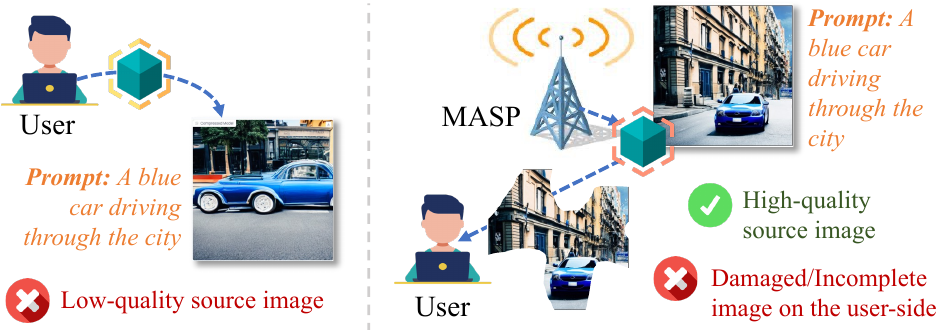}}
\caption{Local AIGC (left) and Traditional Mobile AIGC (right).}
\label{example}
\end{figure}

\section{Motivation and System Model}
\subsection{Motivation}
Without loss of generality, we consider the text-to-image AIGC scenario in this paper.
However, the proposed framework and algorithms are applicable to other forms of AIGC\footnote{To extend the proposed framework to other mobile AIGC scenarios, only the attention generation mechanism needs to be changed, e.g., \cite{VideoAIGC} can be applied in video-oriented AIGC.}.
Suppose that users adopt ``A blue car driving through the city." as the prompt for image generation.
The existing AIGC paradigms include:
\begin{itemize}
    \item \textbf{Local AIGC}: Due to constrained computing resources, if generating images locally, the users can only afford to utilize compressed AIGC models \cite{kim2023architectural}, resulting in poor generation quality [see Fig. \ref{example}(left)]. 
    \item \textbf{Traditional Mobile AIGC}: Leveraging mobile AIGC, the users can call MASPs to generate high-quality images using powerful AIGC models \cite{SDpaper}. Nonetheless, repeated image downloads from the MASPs consume considerable communication resources. Moreover, given the limited wireless bandwidth, large output images may not be fully transmitted. Damaged or incomplete images are useless to users [see Fig. \ref{example}(right)].
\end{itemize}
To this end, we develop G-SemCom for mobile AIGC.
Our goal is to enable mobile users to acquire high-quality images under computing and communication resource constraints.
\renewcommand{\arraystretch}{1.2}
\begin{table}[tpb]
\begin{tabular}{l|p{6.5cm}}
\Xhline{2.2pt}
\rowcolor[rgb]{0.92,0.92,0.92}
\textbf{Notation}& \textbf{Description} \\
\hline
$A_{z}^{\mathbb{R}^{+}}[x, y]$& Cross-model attention map\\
\hline
$A_{z}^{{0, 1}}[x, y]$& Binary attention map\\
\hline
$\beta_t$& Noise added in forward diffusion at step $t$\\
\hline
$\boldsymbol{C}$, $\boldsymbol{C}^{*}$& Boolean dependency matrix (original and compressed)\\
\hline
$\phi_i$ & Channel gain \\
\hline
$\gamma$& Discount factor of ADD\\
\hline
$\boldsymbol{D}$, $\boldsymbol{D}^{*}$& Dependency level matrix (original and compressed)\\
\hline
$F_{t}^{(i)\texttt{d}}$, $F_{t}^{(i)\texttt{u}}$& Cross-modal attention scores for down/upstream blocks \\
\hline
$\boldsymbol{I}$& Identity matrix\\
\hline
$\eta$& Learning rate of ADD\\
\hline
$L_i$& Transmission latency of user $U_i$\\
\hline
$N_0$& Noise power spectral density\\
\hline
$O$& Bandwidth consumption of transmitting each token in $\mathcal{S}^{}{A_i^{{0, 1}}}$\\
\hline
$P$& Transmission power of MASP \\
\hline
$\mathbf{p}$& Prompt provide by user for image generation \\
\hline
$Q_\mathrm{th}$& User threshold for aesthetic quality\\
\hline
$T$ & Number of denoising steps in source image generation and ADD\\
\hline
$\xi$& Threshold for constructing binary attention maps\\
\hline
$\mathbf{x}$& Latent vector used in the diffusion process\\
\hline
$\mathbf{w}$& Word embeddings \\
\hline
$W$& Bandwidth of each RB \\
\hline
$\omega_0, \omega_1, \omega_2$& Weighting factors in optimization problems\\
\hline
$\boldsymbol{S}_i$& Semantic information for user $U_i$\\
\hline
$\mathcal{S}^{}_{A_i^{{0, 1}}}$& Image segments\\
\hline
$\mathbf{s}$& Semantic importance of each prompt word\\
\Xhline{2.2pt}
\end{tabular}
\caption{The main mathematical notations.}
\vspace{-0.3cm}
\end{table}
\renewcommand{\arraystretch}{1}

\subsection{System Model}
\begin{figure}[tbp]
\centerline{\includegraphics[width=0.97\columnwidth]{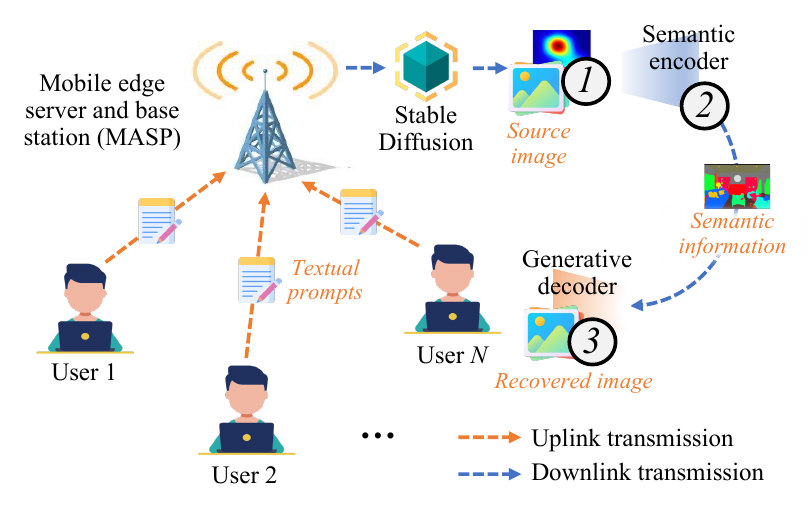}}
\caption{The system model. Step 1: \textit{Cross-modal attention map generation}, Step 2: \textit{Attention-aware semantic extraction}, and Step 3: \textit{Generative decoding}.}
\label{framework}
\end{figure}

As shown in Fig. \ref{framework}, we consider the system with one MASP and $N$ users, denoted by $\mathcal{U}$ = $\{U_1, U_2, \dots, U_N\}$.
However, the model can be extended straightforwardly for multiple MASPs.
To acquire high-quality images, the users first send their prompts to the MASP.
Serving by mobile edge servers and base stations, the MASP operates Stable Diffusion \cite{SDpaper}, the start-of-the-art text-to-image model, and provides generative inference services for the users.
In this way, high-quality images can be generated\footnote{In this paper, the images generated by the MASP are called \textbf{source images} given their role as the semantic source in the G-SemCom framework.}.
Then, G-SemCom is applied to overcome the bandwidth constraints.
Specifically, the MASP generates a series of cross-modal attention maps during the inference, which associate each prompt word with certain source image pixels (Step 1).
After attention-aware semantic extraction (Step 2), only the most semantically important pixels serve as semantic information and are transmitted over a wireless channel.
Then, the users employ a generative decoder, taking semantic information as the prompt to recover the source image (Step 3).
Particularly, to strike an optimal balance between the limited bandwidth and the human-perceptual G-SemCom experience, we formulate a joint semantic encoding and prompt engineering problem.
This aims to optimize the bandwidth allocation among multiple mobile AIGC users. 
Next, we illustrate the transmission model.
Afterward, Sections IV and V discuss the G-SemCom design and the joint optimization problem, respectively.

\subsection{Transmission Model}
We utilize the orthogonal frequency division multiple access (OFDMA) technique \cite{Yining} to model the wireless transmission between the MASP and users.
Specifically, each user is allocated one downlink orthogonal resource block (RB).
Suppose that the $i^{th}$ RB is assigned for transmitting semantic information $\boldsymbol{S}_i$ to user $U_i$, the corresponding downlink channel capacity is defined as \cite{Yining}
\begin{equation}
    c_{i}= W \log _{2}\left(1+\frac{P \phi_{i}}{I_{n}+W N_{0}}\right), i \in \{1, 2, \dots, N\}
\end{equation}
where $W$ is the bandwidth of each RB; $P$ means the transmission power of the MASP, $I_q$ means the interference caused by the base stations that are located in other service areas and use the $i^{th}$ RB, and $N_0$ is the noise power spectral density.
$\phi_{i} = \gamma_{i} \,d_{i}^{-2}$ represents the channel gain between the MASP and user $U_i$ with $\gamma_{i}$ being the Rayleigh fading parameter and $d_{i}$ being their physical distance. 
Here, we consider that the transmission latency between the MASP and user $U_i$ is limited to $L_i$. 
Hence, given the data rate $c_{i}$, the maximum size of semantic information can be determined.
Note that the important mathematical notations used in this paper are summarized in TABLE I.

\section{Cross-Modal G-SemCom for Mobile AIGC}
In this section, we illustrate the design of our G-SemCom framework for mobile AIGC.
First, we introduce the process of source image generation.
Then, we demonstrate the generation of cross-modal attention maps.
Finally, we develop the G-SemCom encoder and decoder.
\begin{figure*}[tbp]
\centerline{\includegraphics[width=1.9\columnwidth]{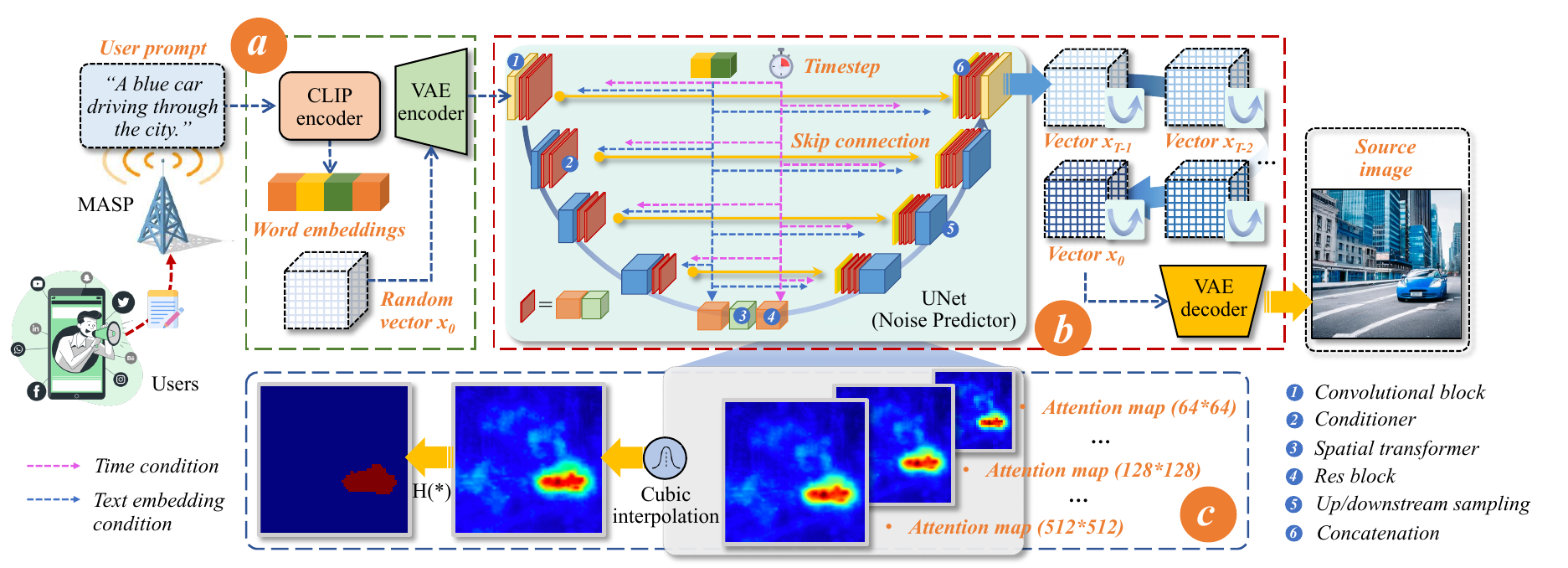}}
\caption{The illustration of generating source images and cross-modal attention maps. (a): The CLIP and VAE modules. (b): The UNet architecture and diffusion process. (c): The attention map of word \texttt{[car]}.}
\label{generation}
\end{figure*}

\subsection{Source Image Generation}
According to the textual prompt, the MASP can generate a source image using Stable Diffusion, depicting the objects and scenes described by the user.
As shown in Figs. \ref{generation}(a) and (b), to realize such text-to-image generations, Stable Diffusion adopts a modular architecture with three components, namely a deep visual-language model called CLIP \cite{CLIP}, a variational autoencoder (VAE) \cite{liu2023optimizing}, and a UNet-based noise predictor \cite{StableUNet}.
To train Stable Diffusion, a large dataset containing massive caption-image pairs is first prepared.
During each training iteration, the fetched image and its caption are encoded by VAE and CLIP into a latent vector $\mathbf{x}_0$ and word embeddings $\mathbf{w} := [w_1, \dots, w_{l_{W}}]$, respectively.
Afterward, a Markov process called forward diffusion is performed.
Specifically, $\mathbf{x}_0$ is gradually perturbed by adding noise for $T$ times, until it becomes a pure Gaussian noise $\mathbf{x}_T$, i.e.,
\begin{equation}
    q\left(\mathbf{x}_{1: T} \mid \mathbf{x}_{0}\right):=\prod_{t=1}^{T} q\left(\mathbf{x}_{t} \mid \mathbf{x}_{t-1}\right),
\end{equation}
where each denoising step satisfies
\begin{equation}
    q\left(\mathbf{x}_{t} \mid \mathbf{x}_{t-1}\right):=\mathcal{N}\left(\mathbf{x}_{t} ; \sqrt{1-\beta_{t}} \mathbf{x}_{t-1}, \beta_{t} \boldsymbol{I}\right),
\end{equation}
where $\boldsymbol{I}$ represents the identity matrix.
Note that $\{\beta_t\}^{T}_{t=1}$ follows a pre-defined schedule so that $p(\mathbf{x}_{T})$ is approximately zero-mean isotropic \cite{DAAM}.
The forward diffusion aims to train the noise predictor, which utilizes the UNet to learn the amount of noise that should be added in each step. 
For generating new images, Stable Diffusion first randomly generates a latent vector $\mathbf{x}_T$.
Then, it performs the reverse diffusion process to subtract noise from $\mathbf{x}_T$.
According to \cite{DDPM}, such a denoising process can be expressed as
\begin{subequations}
\begin{align}
    p_{\theta}&\left(\mathbf{x}_{t-1} \mid \mathbf{x}_{t}\right) := \mathcal{N}\left(\mathbf{x}_{t-1} ; \mu_\theta(\mathbf{x}_t, t, \mathbf{w}), \beta_{t} \boldsymbol{I}\right), \tag{4} \\
    \mu_\theta&(\mathbf{x}_t, t, \mathbf{w}) = \frac{1}{\sqrt{\alpha_{t}}}\left(\mathbf{x}_{t}-\frac{\beta_{t}}{\sqrt{1-\bar{\alpha}_{t}}} \epsilon_{\theta}\left(\mathbf{x}_{t}, t\right)\right), \\
    &\;\;\;\;\;\;\;\;\alpha_t := 1 - \beta_t,\; \bar{\alpha_t} := \prod_{i=1}^{t} \alpha_i,
\end{align}
\end{subequations}
where $\epsilon_{\theta}\left(\mathbf{x}_{t}, t; \mathbf{w}\right)$ means the noise predicted by UNet, and $\theta$ represents the well-trained U-net parameters.
Iteratively processing Eq. (4), the latent representation of the required image, i.e., $\mathbf{x}_0$, can be generated.
Finally, $\mathbf{x}_0$ is decoded by VAE and becomes a high-quality and user-perceivable source image.

\subsection{Cross-Modal Attention Map}
From Eq. (4), we can observe that text embeddings $\mathbf{w}$ condition the image generation, which explains why the generated images are semantically equivalent to user prompts.
As shown in Fig. \ref{generation}(b), in Stable Diffusion, the text embeddings and latent image vector are bridged by UNet's spatial transformer blocks in the form of cross-modal attention. 
To be specific, UNet is basically composed of $K$ downsampling convolutional blocks and the corresponding upsampling blocks [see Fig. \ref{generation}(b)]. 
Suppose that given a latent image vector $\mathbf{x}_t \in \mathbb{R}^{\omega \times h}$ ($t \in \{1, 2, \dots, T\}$), firstly, the downsampling blocks output a series of vectors $\{\mathbf{v}^{\texttt{d}}_{i, t}\}_{i=1}^K$, where $\mathbf{v}^{\texttt{d}}_{i, t} \in \mathbb{R}^{\lceil \frac{\omega}{c_i} \rceil \times \lceil \frac{h}{c_i} \rceil}$ for some $c \, \textgreater \, 1$.
Then, the upsampling blocks iteratively upscale $\mathbf{v}^{\texttt{d}}_{K, t}$ to $\{\mathbf{v}^{\texttt{u}}_{i, t}\}_{i=K-1}^0$, where $\mathbf{v}^{\texttt{u}}_{i, t} \in \mathbb{R}^{\lceil \frac{\omega}{c_i} \rceil \times \lceil \frac{h}{c_i} \rceil}$.
To support conditioned content generation, diffusion-based AIGC models like Stable Diffusion attach two more blocks to each downsampling/upsampling block, namely a resblock and a spatial transformer \cite{StableUNet}. 
They are responsible for providing $t$ and $\mathbf{w}$ conditions in Eq. (4), respectively.
In this paper, we focus on the latter since the cross-modal attention reflects the modality transfer happening during the AIGC inference.
For each downsampling/upsampling block, the cross-modal attention corresponding to it can be denoted by
\begin{equation}
    \displaystyle \mathbf{v}_{i, t}^{\texttt{d}}:=F_{t}^{(i)\texttt{d}}\Big(\hat{\mathbf{v}}_{i, t}^{\texttt{d}}, \mathbf{w}\Big)\left(\boldsymbol{W}_{v}^{(i)} \mathbf{w}\right),
\end{equation}
\begin{equation}
    \displaystyle F_{t}^{(i)\texttt{d}}\!\left(\hat{\mathbf{v}}_{i, t}^{\texttt{d}}, \mathbf{w}\right)\!:=\operatorname{softmax}\!\left(\frac{\left(\boldsymbol{W}_{q}^{(i)} \hat{\mathbf{v}}_{i, t}^{\texttt{d}}\right)\!\left(\boldsymbol{W}_{k}^{(i)} \mathbf{w}\right)\!^{T}}{\sqrt{d}}\right)\!\!,
\end{equation}
where $F_{t}^{(i)\texttt{d}}$ denotes the normalized downsampling attention score array.
The attention of each word $w_z, z \in \{1, 2, \dots, l_{W}\}$ on the 2D intermediate coordinate of the $l^{th}$ head ($l \in \{1, 2, \dots, l_{H}\}$) belonging to the $i^{th}$ downsampling block can be measured with a score within [0, 1].
$\boldsymbol{W}_k$, $\boldsymbol{W}_q$, and $\boldsymbol{W}_v$ are projection matrices with $l_H$ attention heads; $d$ is a scaling factor.
Note that for simplicity, we do not show the equations of upsampling attention score array $F_{t}^{(i)\texttt{u}}$, which are similar to Eqs. (5) and (6).
As shown in Fig. \ref{generation}(c), the intermediate coordinate, in the form of $[x, y]$, is locally mapped to a surrounding affected square area in the source image.
In this way, we can quantify the correlation between the given prompt word and each image pixel according to the attention score value.
However, the downsampling/upsampling blocks of UNet vary in size, resulting in a series of attention maps with different scales.
Based on \cite{DAAM}, as shown in Fig. \ref{generation}(c), we upscale all $F_{t}^{(i)\texttt{d}}$ and $F_{t}^{(i)\texttt{u}}$ to the original image size, i.e., $\omega \times h$, using bicubic interpolation.
Then, the attention scores are summed up over the heads, layers, and diffusion steps, forming the cross-modal attention map as follows
\begin{equation}
    A_{z}^{\mathbb{R}^{+}}[x, y]:=\!\!\sum_{t=1}^{T}\sum_{i=1}^{K}\sum_{\substack{\ell=1}}^{\substack{l_H}} \left(F_{t, (z, \ell)}^{(i) \texttt{d}}[x, y]+F_{t, (z, \ell)}^{(i) \texttt{u}}[x, y]\right),
\end{equation}
where $z$ and $l$ represent the indexes of the word embedding and downsampling/upsampling block, respectively.
$\mathbb{R}^{+}$ indicates that any $A_{z}^{\mathbb{R}^{+}}[x, y]$ belongs to positive real number.
Finally, we generate the binary cross-modal attention maps by 
\begin{equation}
    A_{z}^{\{0, 1\}}[x, y] := \operatorname{H}\left(A_{z}^{\mathbb{R}^{+}}[x, y] \geq \xi \max_{x, y} A_{z}^{\mathbb{R}^{+}}[x, y] \right),
\end{equation}
where $\xi \max_{x, y} A_{z}^{\mathbb{R}^{+}}[x, y]$ is the pre-defined threshold.
$\operatorname{H}(\cdot)$ represents the Heaviside step function, which outputs \textit{\textbf{1}} when the value of $A_{z}^{\{0, 1\}}[x, y]$ exceeds the threshold, and \textit{\textbf{0}} otherwise.
Compared with fine-grained attention scores, i.e., $A_{z}^{\mathbb{R}^{+}}[x, y]$, $A_{z}^{\{0, 1\}}[x, y]$ facilitates the set operations on multiple attention maps, which are discussed below.

\subsection{Attention-Aware Semantic Extraction}
With the cross-modal attention maps, in this part, we extract semantic features from the source image.
The entire procedure is shown in \textbf{Algorithm 1}.
\renewcommand{\arraystretch}{1.2}
\begin{table}
\begin{tabular}{l|p{1.8cm}<{\centering}|p{4.8cm}<{\centering}}
\Xhline{2.2pt}
\rowcolor[rgb]{0.92,0.92,0.92}
\textbf{Type}&\textbf{Definition}&\textbf{Examples}\\
\hline
\textbf{\textit{NN}}&noun& \texttt{dog, people, city} $\dots$\\
\hline
\textbf{\textit{PROPN}}&proper noun& \texttt{iPhone, IEEE, Alice} $\dots$\\
\hline
\textbf{\textit{NUM}}&numeral& \texttt{one, two, 100, 10th} $\dots$\\
\hline
\textbf{\textit{ADJ}}&adjective& \texttt{red, beautiful, good} $\dots$\\
\hline
\textbf{\textit{VERB}}&actions&\texttt{run, drive, think} $\dots$\\
\hline
\textbf{\textit{ADV}}&adverb&\texttt{quickly, rapidly, here} $\dots$\\
\hline
\textbf{\textit{ADP}}&adposition&\texttt{on, at, through} $\dots$\\
\hline
\textbf{\textit{X}}&other words/symbols& \texttt{(a, the), (and, but), (Wow), (she, he), (".", "\$")} $\dots$\\
\Xhline{2.2pt}
\end{tabular}
\caption{The Part-of-Speech of textual prompts. Note that \textbf{\textit{X}} includes all the words/symbols that do not belong to any of the above types.}
\vspace{-0.3cm}
\end{table}
\renewcommand{\arraystretch}{1}

\subsubsection{Textual Prompt Deconstruction}
\begin{figure}[tbp]
\centerline{\includegraphics[width=0.99\columnwidth]{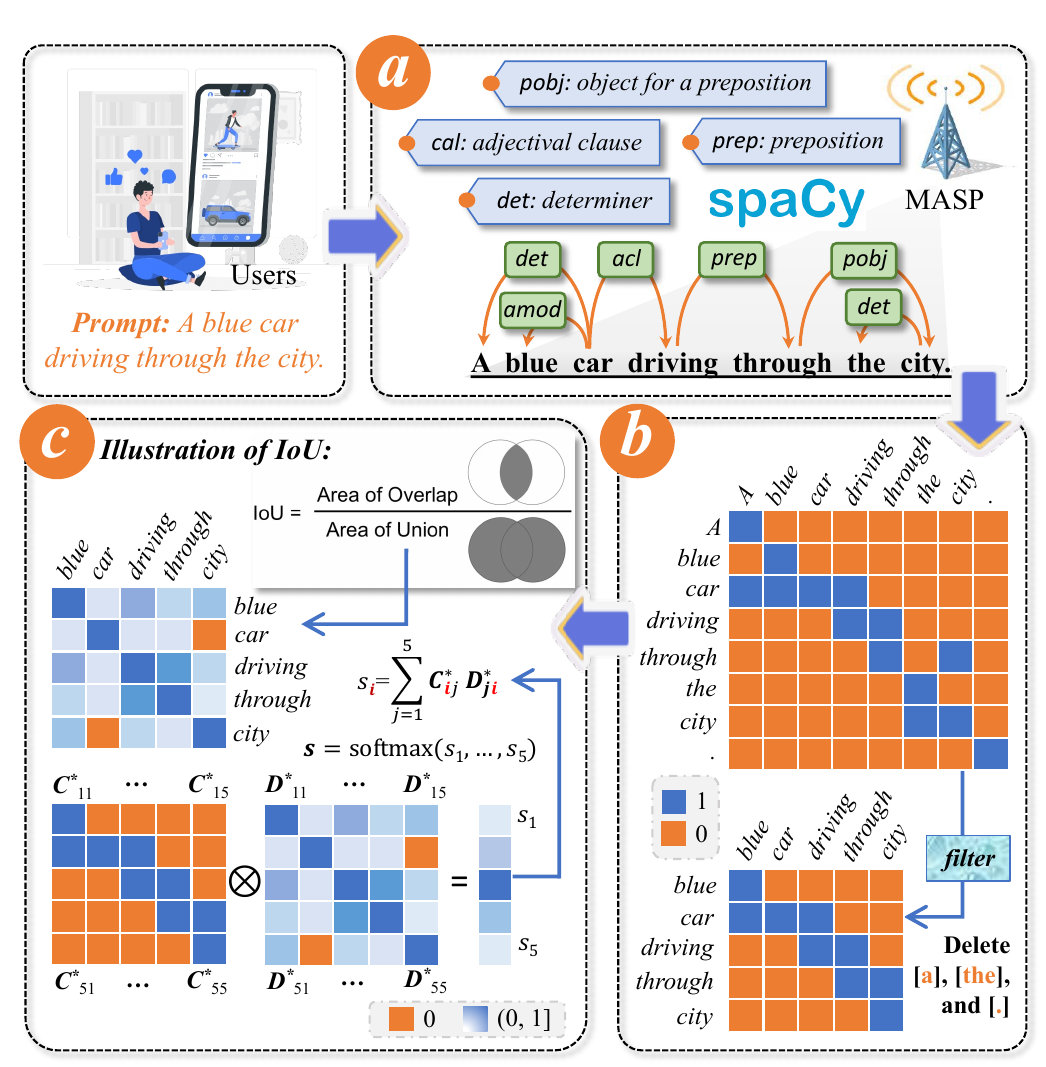}}
\caption{Attention-aware semantic extraction. (a): Dependency parsing. (b): Boolean dependency matrices $\boldsymbol{C}$ and $\boldsymbol{C}^{*}$. (c): Dependency level matrix $\boldsymbol{D}^{*}$ and $\otimes$ operation.}
\label{extraction}
\end{figure}

Firstly, the MASP analyzes the textual prompts provided by users, denoted by $\mathbf{p}\,:= [p_1, p_2, \dots, p_M]$, trying to understand the semantic meaning of users' requirements.
To do so, we utilize \textit{Spacy} \cite{DAAM} to perform the \textit{Part-of-Speech} tagging, i.e., classifying the words according to their linguistic functions.
As shown in TABLE II, we consider seven part-of-speech types that are semantically important \cite{DAAM} and use $\textit{\textbf{X}}$ to include all other types (e.g., determiner, interjection, and conjunction) with less semantic meanings.
Take ``A blue car driving through the city." as an example.
Words \texttt{[car]} and \texttt{[city]} belong to $\textit{\textbf{NN}}$; \texttt{[blue]} belongs to $\textit{\textbf{ADJ}}$; \texttt{[driving]} belongs to $\textit{\textbf{VERB}}$; \texttt{[through]} belongs to \textit{\textbf{ADP}}; \texttt{[A]}, \texttt{[the]}, and \texttt{[.]} belong to $\textit{\textbf{X}}$.
Afterward, we perform dependency parsing \cite{NLPDependency}, aiming to analyze the grammatical structure of $\mathbf{p}$ and find out related words as well as their correlation.
As shown in Fig. \ref{extraction}(a), each dependency item takes the form of an arrow, from head to the word that modifies it, called dependent.
Similarly, this step can be realized by \textit{Spacy}.
In this way, the Boolean dependency matrix $\boldsymbol{C} \in \mathbb{R}^{M\times M}$ can be constructed as follows
\begin{equation}
    \boldsymbol{C} = \left[\begin{array}{ccccc}
R_{(1\leftarrow 1)}^{\{0,\,1\}} \!\!& \cdots &\!\! R_{(1\leftarrow i)}^{\{0,\,1\}} \!\!& \cdots &\!\! R_{(1\leftarrow M)}^{\{0,\,1\}} \\
R_{(i\leftarrow 1)}^{\{0,\,1\}} \!\!& \ddots &\!\! R_{(i\leftarrow i)}^{\{0,\,1\}} \!\!& \ddots &\!\! R_{(i\leftarrow M)}^{\{0,\,1\}} \\
R_{(M\leftarrow 1)}^{\{0,\,1\}} \!\!& \cdots &\!\! R_{(M\leftarrow i)}^{\{0,\,1\}} \!\!& \cdots &\!\! R_{(M\leftarrow M)}^{\{0,\,1\}}
\end{array}\right],
\end{equation}
where $R_{(i\leftarrow j)}^{\{0,1\}}$ ($i, j \in \{1, 2, \dots, M\}$) takes the value \textbf{\textit{1}} if the dependency exists between $p_i$ and $p_j$, i.e., the head and the dependent, respectively, and \textbf{\textit{0}} otherwise.

Due to weak semantic meaning, the words belonging to $\textit{\textbf{X}}$ can be filtered out to reduce the computation complexity.
Accordingly, $\boldsymbol{C}$ can be compressed to $\boldsymbol{C}^{*} \in \mathbb{R}^{(M-\zeta) \times (M-\zeta)}$, where $\zeta$ represents the number of $\textit{\textbf{X}}$-type words in $\mathbf{p}$.
The original and compressed dependency matrices of ``A blue car driving through the city." are shown in Fig. \ref{extraction}(b).

$\boldsymbol{C}^{*}$ can reflect the importance of each word based on the number of dependencies that it involves.
Nonetheless, according to the types of the head and the dependent, as well as the function that the dependent acts on the head, there exist more than 18 kinds of dependencies \cite{NLPDependency}.
Some dependencies, such as (\texttt{amod}:\,\texttt{[blue]}\,$\leftarrow$\,\texttt{[car]}), are strong, while the others, such as (\texttt{det}:\,\texttt{[A]}\,$\leftarrow$\,\texttt{[car]}), are weak.
To this end, we leverage the \textit{Mean Intersection over Union} (mIoU) to calculate the fine-grained semantic importance of each word.
Suppose that the pixels of the entire source image construct the Universe $\mathcal{S}$, and the pixels included by the binary attention maps of words $p_i$ and $p_j$ are sets $\mathcal{S}_{A_i^{\{0, 1\}}}$ and $\mathcal{S}_{A_j^{\{0, 1\}}}$, respectively. mIoU can be derived as 
\begin{equation}
    \mathrm{mIoU}_{(i\leftarrow j)} = \frac{|\mathcal{S}_{A_i^{\{0, 1\}}} \cap \mathcal{S}_{A_j^{\{0, 1\}}}|}{|\mathcal{S}_{A_i^{\{0, 1\}}} \cup \mathcal{S}_{A_j^{\{0, 1\}}}|}.
\end{equation}
As shown in Fig. \ref{extraction}(c), $\mathrm{mIoU}_{(i\leftarrow j)}$ can measure the similarity of the areas covered by the binary attention maps of words $p_i$ and $p_j$ and is within [0, 1].
Consequently, the higher the mIoU value, the stronger the dependency exists in two words.
Using mIoU, we can acquire the following dependency level matrix, denoted by $\boldsymbol{D}^{*} \in \mathbb{R}^{(M-\zeta) \times (M-\zeta)}$.
\begin{equation}
    \boldsymbol{D}^{*} = \left[\begin{array}{ccccc}
L_{(1\leftarrow 1)}^{\{0, \mathbb{R}^{+}\}} \!\!\!& \cdots &\!\!\! L_{(1\leftarrow i)}^{\{0, \mathbb{R}^{+}\}} \!\!\!& \cdots &\!\!\! L_{(1\leftarrow \sigma)}^{\{0, \mathbb{R}^{+}\}} \\
L_{(i\leftarrow 1)}^{\{0, \mathbb{R}^{+}\}} \!\!\!& \ddots &\!\!\! L_{(i\leftarrow i)}^{\{0, \mathbb{R}^{+}\}} \!\!\!& \ddots &\!\!\! L_{(i\leftarrow \sigma)}^{\{0, \mathbb{R}^{+}\}} \\
L_{(\sigma\leftarrow 1)}^{\{0, \mathbb{R}^{+}\}} \!\!\!& \cdots &\!\!\! L_{(\sigma\leftarrow i)}^{\{0, \mathbb{R}^{+}\}} \!\!\!& \cdots &\!\!\! L_{(\sigma\leftarrow \sigma)}^{\{0, \mathbb{R}^{+}\}}
\end{array}\right],
\end{equation}
where $L_{(i\leftarrow j)}^{\{0, \mathbb{R}^{+}\}}$ ($i, j \in \{1, 2, \dots, (M-\zeta)\}$) takes the value of $\mathrm{mIoU}_{(i\leftarrow j)}$; $\sigma$ equals $M-\zeta$.
Finally, the semantic importance of each word, denoted as $\mathbf{s} := \{s_1, s_2, \dots, s_{M-\zeta}\}$, in which the $\textbf{X}$-type words have been filtered out, can be derived as
\begin{equation}
    \mathbf{s} = \operatorname{softmax}(\boldsymbol{C}^{*} \otimes \boldsymbol{D}^{*}),
\end{equation}
where $\otimes$ represents a self-defined matrix multiplication operation and is shown in Fig. \ref{extraction}(c).
Note that $\mathbf{s}$ is normalized by $\operatorname{softmax}$, ensuring that every $s_i$ ($i \in \{1, 2, \dots, (M-\zeta)\}$) is within [0, 1] and $\sum_{i=1}^{M-\zeta} s_i = 1$.
In our example, the importance of \texttt{[blue]}, \texttt{[car]}, \texttt{[driving]}, \texttt{[through]}, and \texttt{[city]} are 0.16, 0.20, 0.31, 0.17, and 0.16, respectively.
Hence, we can conclude that two major objects and their relationship, i.e., \texttt{[car]}, \texttt{[driving]}, and \texttt{[city]}, cover the major semantic meaning of the entire source image.
In contrast, the preposition, i.e., \texttt{[through]}, is weak in terms of semantic importance.
\begin{figure}[tbp]
\centerline{\includegraphics[width=0.95\columnwidth]{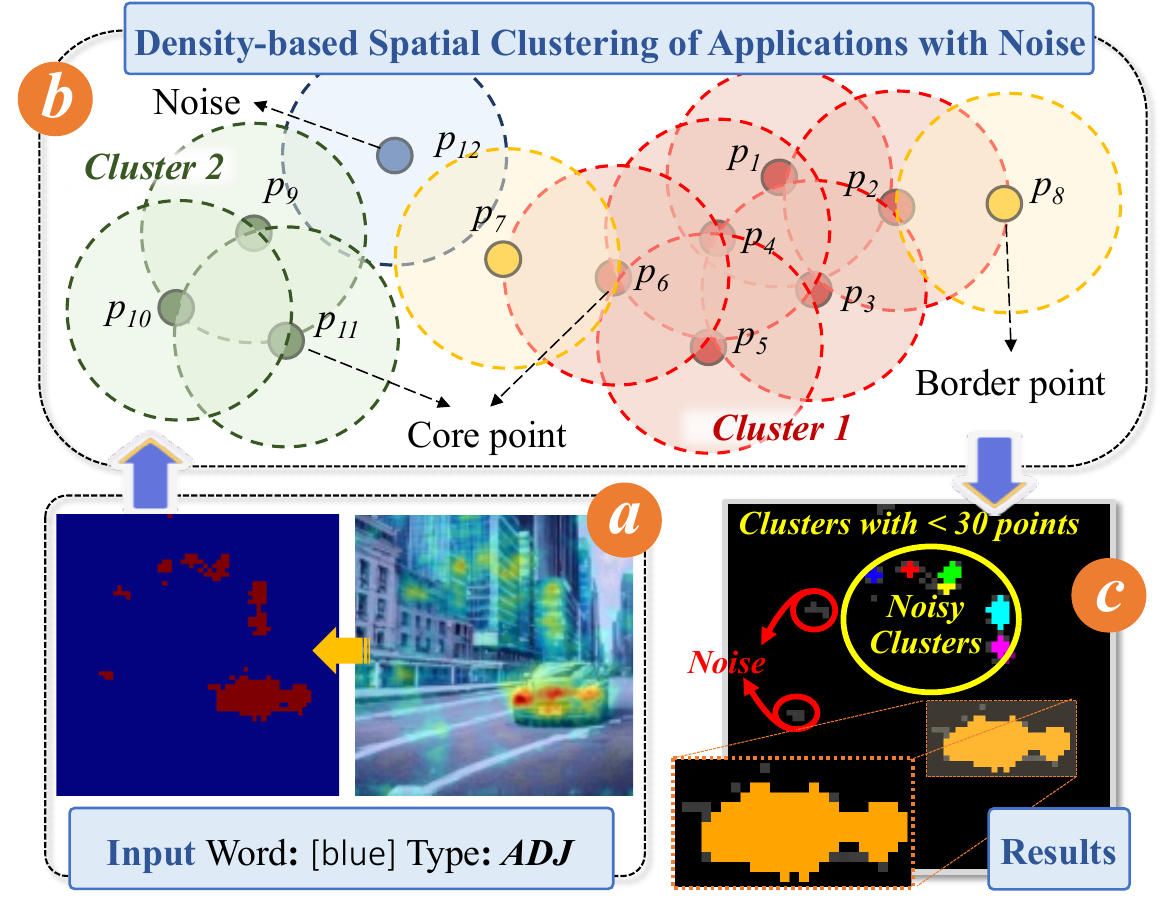}}
\caption{Illustration of attention clustering. (a): The raw and binary attention maps. (b): The illustration of the DBSCAN algorithm. (c): The clustering results. Note that noise and noisy clusters will be filtered out.}
\label{DBSCAN}
\end{figure}

\subsubsection{Visual Prompt Segmentation}
Up till now, we can evaluate the semantic importance of each word and link it to certain areas of the source image.
However, as shown in Fig. \ref{DBSCAN}(a), the attention distribution of some words (especially \textit{\textbf{ADV}-} and \textbf{\textit{VERB}}-type ones) in the source image is scattered, containing a lot of outlier noise.
Such noise not only wastes bandwidth resources but also increases the difficulty of image recovery.
To this end, we intend to perform clustering on the cross-modal attention maps according to the attention density and remove the noise.
Therefore, leveraging \textit{Density-Based Spatial Clustering of Applications with Noise} (DBSCAN) \cite{DBSCAN}, we present the attention-aware visual prompt segmentation.
As shown in \textbf{Algorithm 1}, the proposed method contains the following three steps.
\begin{itemize}
    \item \textbf{Attention Point Clustering}: Given a cross-modal attention map, i.e., $A_{z}^{\{0, 1\}}$ ($z \in \{1, 2, \dots, (M-\zeta)\}$), we first cluster dense attention points and filter the noise, using a sophisticated clustering algorithm called DBSCAN\footnote{The DBSCAN is applied due to its wide adoption and well-proven performance in clustering points following complex distributions. Other clustering algorithms, such as k-means and OPTICS, can also be applied.}. DBSCAN conducts clustering by first detecting all the core points that have at least $\Omega$ neighbors, i.e.,
    \begin{equation}
    N_{\varepsilon}(p) \geq \Omega, N_{\varepsilon}(p)=|\{q \in \mathcal{S}_{A_z^{\{0, 1\}}} \!\mid\! \operatorname{d}(p, q) \leq \varepsilon\}|,
    \end{equation}
    where $\varepsilon$ and $\Omega$ are user-defined and represent the distance threshold and the required number of neighbors within $\varepsilon$, respectively. $\operatorname{d}$ is the function for distance measurement. As shown in Fig. \ref{DBSCAN}(b), DBSCAN starts from a random core point, e.g., $p_1$, and iteratively groups all neighboring core points (i.e., $p_2 \dots, p_6$) into the same cluster. The border points that are close to any of the aforementioned core points are also included (i.e., $p_7$ and $p_8$). Afterward, another core point that has not been clustered, e.g., $p_8$, can be selected, and the above process is repeated. The algorithm will stop when all the core points are clustered. Accordingly, the remaining points are viewed as noise. More details are shown in \textbf{Algorithm 1}. Fig. \ref{DBSCAN}(c) illustrates the clustering results of the attention map of the word \texttt{[blue]}. We can observe that those scattered attention points that do not have clear semantic meanings are identified as noise.
    \item \textbf{Source Image Segmentation}: Step 1 will be performed for all the acquired attention maps. Afterward, we can acquire a series of clean image segments, denoted by $\mathcal{S}^{*}_{A_z^{\{0, 1\}}}$ ($\forall z \in \{1, 2, \dots, (M-\zeta)\}$).
    \item \textbf{Semantic Information Packing}: Finally, the MASP packs semantic information, which guides the users to recover semantically similar and high-quality images. Therefore, the pixels owning stronger semantic meanings should be prioritized. Next, we reorder $\mathbf{p}$ according to the semantic importance of each word, i.e., $\mathbf{s}$. Then, the semantic information matrix $\boldsymbol{S}$ can be generated.
    \begin{equation}
        \boldsymbol{S} = \left[\begin{array}{l}
        \mathbf{s}_1 = \mathcal{S}^{*}_{A_1^{\{0, 1\}}}, \\
        \mathbf{s}_2 = \mathcal{S}^{*}_{A_2^{\{0, 1\}}} \setminus \mathbf{s}_1, \\
        \mathbf{s}_3 = \mathcal{S}^{*}_{A_3^{\{0, 1\}}} \setminus (\mathbf{s}_1 \cup \mathbf{s}_2), \\
        \dots \\
        \mathbf{s}_{M-\zeta} = \mathcal{S}^{*}_{A_{M-\zeta}^{\{0, 1\}}} \setminus (\mathbf{s}_1 \!\cup \!\dots \!\cup \mathbf{s}_{M-\zeta-1}) \\
        \end{array}\right]\!.
    \end{equation}
    The MASP will send $[\mathbf{s}_1, \mathbf{s}_2, \dots, \mathbf{s}_{M-\zeta}]$ in sequence until all available bandwidth is used.
\end{itemize}
\begin{algorithm}[tpb]
\footnotesize \caption{The Operations on the MASP-Side}
\begin{algorithmic}[1]
\Require  
$g_0$, $\mathbf{p} = {[p_1, p_2, \dots, p_M]}$, $A_{z}^{\{0, 1\}}[x, y], z \in \{1, 2, \dots, M\}$ \textit{\#\#\, source image, prompt, and binary attention maps}
\Ensure 
$\boldsymbol{I}$ \textit{\#\#\, semantic information}
\Procedure{Textual Prompt Extraction}{$\mathbf{p}$} 
\State Call \textit{spacy} to perform text-to-speech tagging and dependency parsing
\State Initialize $\boldsymbol{C}$ = $0^{M\times M}$
\ForAll{$p_i \in \mathbf{p}$ } \textit{\#\# row iteration}
\ForAll{$p_j \in \mathbf{p}$ } \textit{\#\# column iteration}
\If{$i = j$}
\State $\boldsymbol{C}_{ij}$ = 1 \textit{\#\# each world is correlated to itself}
\Else
\If{$p_i$ and $p_j$ has dependency with $p_i$ as the head and $p_j$ as the dependent}
\State $\boldsymbol{C}_{ij}$ = 1 \textit{\#\# mark the dependency}
\EndIf
\EndIf
\EndFor
\EndFor
\ForAll{$p_i \in \mathbf{p}$} \textit{\#\# filter non-important words, acquiring $\boldsymbol{C}^{*}$}
\If{$p_i$ belongs to \textit{\textbf{X}}-type}
\State Delete the $i^{th}$ column and row of $\boldsymbol{C}$
\EndIf
\EndFor
\State Initialize $\boldsymbol{D}^{*} = \boldsymbol{C}^{*}$ \textit{\#\# the compressed dependency matrix}
\State Initialize $\boldsymbol{D}^{*}$ = $0^{(M-\zeta)\times (M-\zeta)}$
\ForAll{$\boldsymbol{D}^{*}_{ij} \in \boldsymbol{D}^{*}$}
\State $\boldsymbol{D}^{*}_{ij} = \mathrm{IoU}_{(i \leftarrow j)}$
\EndFor
\State $\mathbf{s} = \operatorname{softmax}(\boldsymbol{C}^{*} \otimes \boldsymbol{D}^{*})$
\EndProcedure
\Statex
\Procedure{Visual Prompt Segmentation}{$A_{z}^{\{0, 1\}}[x, y]$}
\ForAll{$z \in \{1, 2, \dots, N\}$}
\State Call DBSCAN to cluster the attention points of $A_{z}^{\{0, 1\}}[x, y]$
\State Filter noise points, as well as the clusters with less than 30 points
\EndFor
\EndProcedure
\Statex
\Procedure{Semantic Information Packing}{$A_{z}^{\{0, 1\}}[x, y]$, $g_0$, $\mathbf{p}$}
\ForAll{$A_{z}^{\{0, 1\}}[x, y], z \in \{1, 2, \dots, N\}$}
\State Sort the attention maps according to $\mathbf{s}$
\EndFor
\State Construct $\boldsymbol{S}$ according to Eq. (14)
\State Send as many tokens in $\boldsymbol{S}$ as possible
\EndProcedure
\end{algorithmic}
\end{algorithm}

\subsubsection{Discussion}
By generating cross-modal attention maps, analyzing the semantic meaning of user prompts, performing attention-aware segmentation of the source image, and only transmitting the semantically important pixels, the size of data that users should download can be efficiently reduced.
Furthermore, the proposed G-SemCom framework will not incur considerable workloads for the MASP.
Firstly, the cross-modal attention maps can be regarded as by-products during the source image generation, causing no additional computation costs.
The computation complexity of DBSCAN is $\mathcal{O}(n\,\operatorname{log} n)$, where $n$ represents the number of attention points \cite{DBSCANComp}.
The remaining operations, such as matrix filtering and multiplications, have the complexity of $\mathcal{O}\left(M-\zeta\right)$ to $\mathcal{O}\left((M-\zeta)^2\right)$.
Despite exponential complexity, the practical computation overhead can be ignored due to the small word numbers of user prompts (typically 10-100).
Therefore, we can conclude that the proposed mechanisms will not affect the performance of the MASP for performing AIGC inferences, which is resource-intensive and occupies most of the computing resources.
More experimental results and analysis are described in Section VI.

\subsection{Generative Semantic Decoder}
After the MASP finishes textual prompt deconstruction and visual prompt segmentation, it can send well-packed semantic information to users.
The users can recover the source image by inpainting the pixels that are not transmitted.
Note that the un-transmitted parts only have weak semantic importance.
For instance, for the image generated from ``A blue car driving through the city.", the road and sky are not mentioned in the user prompts.
Therefore, the illustration of them will not affect the semantic correctness of the recovered image.
Moreover, from the image composition perspective, the road and sky are only used as background to connect semantically strong objects (i.e., car and city), slightly influencing image aesthetic quality.
Hence, the users can adopt various lightweight open-source image inpainting models based on diffusion or generative adversarial networks.
We can consider that the image recovery model (e.g., \cite{Inpainting1} or \cite{Inpainting2}) is well-trained and shared among all users as the generative semantic decoders.

\section{Joint Semantic Encoding and Prompt Engineering}
To effectively reflect the human-perceptual experience of the G-SemCom-aided mobile AIGC services, this section first designs a metric called Joint Perceptual Similarity and Quality (JPSQ).
Then, we formulate the joint optimization problem and present an Attention-aware Deep Diffusion (ADD) algorithm to solve it.
\begin{figure*}[tbp]
\centerline{\includegraphics[width=1.75\columnwidth]{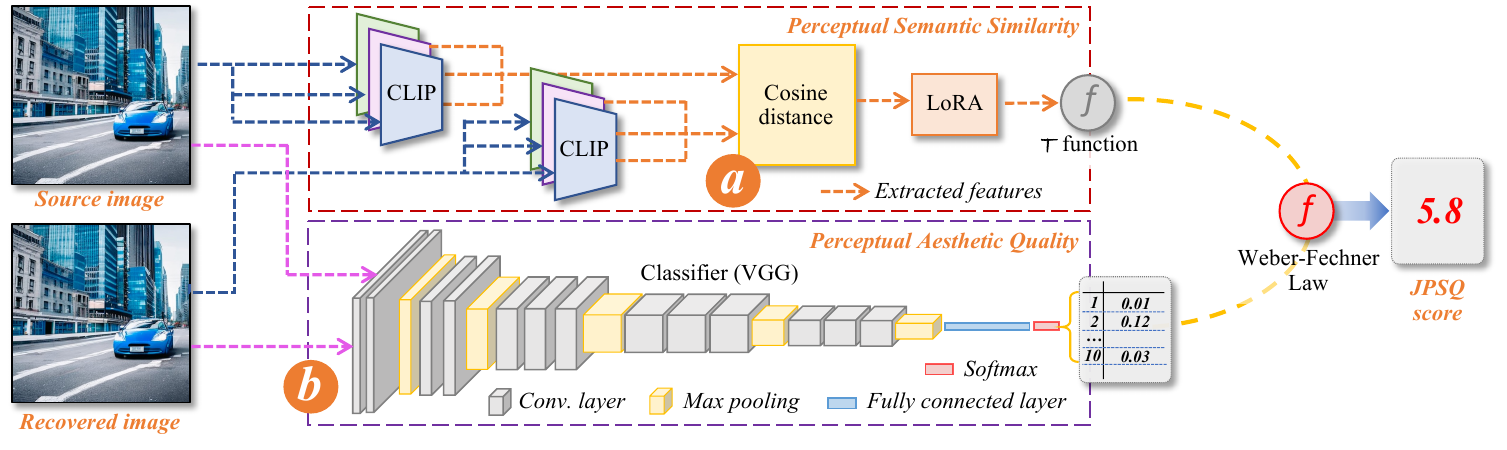}}
\caption{The definition of JPSQ metric for G-SemCom-aided Mobile AIGC, which consists of perceptual semantic similarity and aesthetic quality. \texttt{(a)}: The framework of DreamSim metric. \texttt{(b)}: The framework of NIMA metric. Note that these two metrics are flexible. Hence, the CLIP and VGG 16 can be replaced by other models for different application scenarios.}
\label{JPSQ}
\end{figure*}

\subsection{JPSQ Definition}
Traditionally, to evaluate the effectiveness of SemCom, users can adopt pixel- or structure-level metrics, such as \textit{Mean Square Error} and \textit{Structural Similarity Index Metric}, to measure the similarity between the source and recovered images \cite{SSIM}.
However, these metrics can only capture the difference in terms of luminance, contrast, and structure while failing to consider the image semantics.
Then, task-oriented metrics for SemCom have been presented, emphasizing whether the proposed SemCom framework can accomplish specific communication tasks \cite{Taskoriented, Taskoriented2}.
For instance, the authors in \cite{9796572} utilized the classification accuracy for the observed objects to evaluate the effectiveness of the UAV-based SemCom.
Following this principle, we design a novel task-oriented metric for G-SemCom in mobile AIGC called JPSQ.
Particularly, in G-SemCom-aided mobile AIGC, the semantic information sent by the MASP undertakes two tasks.
First, it guarantees that users can recover images that maintain the same semantics as source images.
Meanwhile, recall that the users aim to acquire high-quality AIGC images.
Hence, the semantic information also serves as the prompts fed to the generative decoder, facilitating it to recover images with high aesthetic quality.
Motivated by this, we jointly consider the semantic similarity and image quality when designing JPSQ.
Furthermore, considering that AI-generated images are consumed by human users, we utilize learning-based metrics trained on large-scale human feedback datasets rather than mathematical methods to capture human perceptual similarity and quality.
Last but not least, we adopt the Weber-Fechner Law \cite{WF_Law} to fuse these two aspects and construct JPSQ.

\subsubsection{Perpetual Semantic Similarity}
To evaluate the perpetual semantic similarity from the user perspective, we utilize the state-of-the-art learning-based metric called \textit{DreamSim} \cite{DreamSim}.
As shown in Fig. \ref{JPSQ}(a), the difference between each pair of images $(g_0, g_1)$ is measured by the cosine distance, i.e.,
\begin{equation}
    D(g_0, g_1; f_\theta) = 1 - \operatorname{cosine}(f_\theta(g_0), f_\theta(g_1)),
\end{equation}
where $f_\theta$ represents the learnable network that extracts important perceptual semantic features from input images.
Such a network is assembled by multiple pretrained models, such as DNIO \cite{DINO} and OpenCLIP \cite{OpenCLIP} and fine-tuned by Low-Rank Adaptation (LoRA) mechanisms, which align the backbone models with the similarity evaluation task.
The smaller the DreamSim value, the more similar the two images are. 

\subsubsection{Perpetual Aesthetic Quality}
To measure the aesthetic quality of the recovered images, we adopt a learning-based image assessment framework called NIMA \cite{NIMA}.
As shown in Fig \ref{JPSQ}(b), NIMA converts the image quality measurement to a classification problem, with ten possible classes representing the quality score from 1 to 10.
Such classification is realized by a pluggable classifier network, supporting VGG16, Inceptio-v2, and MobileNet \cite{NIMA}.
Accordingly, the classification output is defined as $\mathbf{c}(g) = [c_1, c_2, \dots, c_{10}]$, where $c_i$ indicates the probability that the given image $g$ achieves score $i$.
Note that $\sum_{i=1}^{10}c_i = 1$ can be guaranteed since a $\operatorname{softmax}$ operation is employed.
Finally, the aesthetic quality of image $g$ can be expressed as:
\begin{equation}
\begin{split}
  \;\;\;\;\;\;\;\;\;\;\;Q(g) & \sim \mathcal{N}(\mu, \sigma), \\  \mu = \sum_{i=1}^{10} i \times c_i, \;\; \sigma& = \sqrt{\left(\sum_{i=1}^{10}(i - \mu)^2 \times c_i \right)}.
\end{split}
\end{equation}
In this paper, we utilize $\mu$ acquired by NIMA to reflect the aesthetic quality of images.

\subsubsection{Metric Fusion}
Finally, we fuse the perceptual semantic similarity and aesthetic quality by Weber-Fechner Law \cite{WF_Law}.
Denoting source and recovered images as $g_0$ and $g_1$, respectively, JPSQ can be calculated as
\begin{equation}
    \mathcal{J}(g_0, g_1) = \mathcal{T}(D(g_0, g_1; f_\theta))\,\operatorname{ln}\left(\frac{\omega_0Q(g_1)}{Q_{\mathrm{th}}}\right),
\end{equation}
where $\omega_0$ serves as a weighting factor and $Q_{\mathrm{th}}$ indicates the minimal image quality required by users.
$\mathcal{T}$ is defined as
\begin{equation}
    \mathcal{T}(t) = \frac{t_{max} - t}{t_{max} - t_{min}},
\end{equation}
where $t_{min}$ and $t_{max}$ represent the lower and upper bounds of the DreamSim score, respectively, and $t_{min}$ is 0.
In this paper, we acquire $t_{max}$ for our case by generating 1000 AIGC images, measuring their DreamSim scores with a pure Gaussian noise, and calculating the average.
Function $\mathcal{T}(\cdot)$ plays two roles.
First, the denominator inverts the differences reflected by the DreamSim score into similarities.
In addition, the effect of the magnitudes can be eliminated \cite{Law}.

\subsection{Problem Formulation}
In this part, we formulate the joint semantic encoding and prompt engineering problem based on JPSQ.
Recall that in our OFDMA-based transmission model, each user can be assigned an RB for receiving data from the MASP.
However, the overall bandwidth resources of the MASP are limited.
Therefore, we intend to optimize the bandwidth allocation among users, acquiring the best trade-off between the overall G-SemCom performance and consumed bandwidth.
The optimization problem can be formulated as follows:
\begin{subequations}
\begin{align}
\max _{\boldsymbol{b_i}} & \sum_{i=1}^{N} \left[\left(\omega_1\mathcal{J}(g_0^{i}, g_1^{b_i})\cdot\operatorname{H}(Q(g_1^{b_i}) \! \geq \! Q_{\mathrm{th}})-\omega_2b_i\right)\right] \tag{19}\\
\text { s.t. } & 0 \leq b_i \leq \min \{L_ic_i, O|\boldsymbol{S}_i|\}, \forall i \in \{1, 2, \dots, N\} \tag{19a}
\end{align}
\end{subequations}
where $b_i$ means the bandwidth resources allocated to user $U_i$. $g_1^{b_i}$ represents the recovered image using the bandwidth of $b_i$, and $g_0^{i}$ is the corresponding source image. Note that a step function $\operatorname{H}(\cdot)$ is applied since the images whose quality is lower than the user threshold are meaningless in AIGC. Finally, $L_i$ means the communication latency threshold between user $U_i$ and the MASP, and $O$ indicates the bandwidth consumption for transmitting each item in $\boldsymbol{S}_i$.

\subsection{Components of Attention-aware Deep Diffusion}
Traditionally, the joint optimization problem can be solved by various DRL-based methods, such as Proximal Policy Optimization (PPO) and Soft Actor-Critic (SAC) algorithms.
However, since the state in our problem is high-dimensional (i.e., the attention map), the existing methods may lack enough exploration ability and yield only sub-optimal solutions.
Hence, to realize efficient bandwidth allocation in complex environments, we introduce a deep diffusion module into traditional DRL for policy optimization, forming the ADD algorithm.
Next, we demonstrate the components of the proposed ADD algorithm, which is composed of five parts, namely agent, state, action, policy, and reward.
\begin{itemize}
    \item \textbf{Agent}: In the proposed G-SemCom-aided mobile AIGC, the agent represents the MASP, which allocates available bandwidth among multiple users for transmitting semantic information.
    \item \textbf{State}: The state of the mobile AIGC environment takes the form of $\mathbf{s}$ := [$\boldsymbol{S}_1$, $\boldsymbol{S}_2$, $\dots$, $\boldsymbol{S}_N$], i.e., the attention-based semantic information of the source images generated for users $U_1$ to $U_N$. Note that to reduce the complexity of ADD for representing and learning the states, the original \textit{512} $\times$ \textit{512} attention maps are resized to \textit{16} $\times$ \textit{16}. The state space is a discrete space, using Boolean values to indicate whether a certain pixel is associated with the user prompts.
    \item \textbf{Action}: We define an action of the agent in our state as a vector $\mathbf{b} := \{b_1, b_2, \dots, b_N\}$, denoting the bandwidth allocating to each user. With $\mathbf{b}$, the MASP encodes the semantic information for each user $U_i$ ($i \in [1, 2, \dots, N]$), i.e., calculating the number of pixels that can be sent by $b_i/O$ and sending the pixels following the mechanism stated in Section IV-C. Recall that $O$ represents the bandwidth usage for sending each pixel.
    \item \textbf{Policy}: The policy refers to the probability of the agent taking action $\mathbf{b}$ at the state $\mathbf{s}$. Particularly, the ADD algorithm adopts a deep diffusion network parameterized by $\theta$ to learn the relationship between the input state $\mathbf{s}$ and the output action $\mathbf{b}$ that can optimize the reward defined in Eq. (19). Such a policy network can be expressed as $\pi_\theta(\mathbf{s}, \mathbf{b}) = P(\mathbf{b}|\mathbf{s})$.
    \item \textbf{Reward}: Finally, given the environment state $\mathbf{s}$, The reward of taking action $\mathbf{b}$ can be defined as $R(\mathbf{b}|\mathbf{s}) = \sum_{i=1}^{N} \left[\left(\omega_1\mathcal{J}(g_0^{i}, g_1^{b_i})\cdot\operatorname{H}(Q(g_1^{b_i}) \! \geq \! Q_{\mathrm{th}})-\omega_2b_i\right)\right]$. Note that if the constraint is not satisfied, we use a negative reward as the penalty.
\end{itemize}
\begin{algorithm}[tpb]
\footnotesize \caption{The Procedure of ADD Algorithm}
\begin{algorithmic}[1]
\Require  
$\mathbf{s}$, $N_b$, $T$, $\eta$, $\gamma$ \textit{\#\#\, AIGC environment, batch size, diffusion step number, discount factor, and learning rate}
\Ensure 
$\mathbf{b}$ \textit{\#\#\, bandwidth allocation scheme}
\Procedure{ADD Training}{$\mathbf{s}$, $N_b$, $T$, $\eta$, $\gamma$} 
\State Initialize networks: policy generation network $\epsilon_{\theta}$, Q-networks $Q_{v_1}$, $Q_{v_2}$, $Q_{v_1}^{*}$, and $Q_{v_2}^{*}$.
\While{not converged}
\State Initialize random noise $\mathbf{b}_T$; generate bandwidth allocation scheme $\mathbf{b}_0$ by denoising process shown in Eq. (20).
\State Add exploration noise to $\mathbf{b}_0$.
\State Execute resource allocation and calculate utility $u$ by Eq. (19).
\State Store the record ($\mathbf{s}, \mathbf{b}_0, u$) in the replay buffer
\State Randomly select $N_b$ records
\State Update the policy generation network by Eq. (22)
\State Update the Q-networks by Eq. (23)
\EndWhile
\EndProcedure
\Statex
\Procedure{ADD Inference}{$\mathbf{s}$, $N_b$, $T$, $\eta$, $\gamma$}
\State Observe the environment $\mathbf{s}$
\State Generate bandwidth allocation scheme $\mathbf{b}_0$
\State \textbf{Return} $\mathbf{b}_0$
\EndProcedure
\end{algorithmic}
\end{algorithm}

\subsection{Attention-aware Deep Diffusion for Optimization}
The deep diffusion network is introduced to learn the optimized policy $\pi_\theta(\mathbf{s}, \mathbf{b})$~\cite{du2023beyond}.
Following the diffusion principle, the final action $\mathbf{b}_0$ can be generated from random noise $\mathbf{b}_T$ after $T$ steps of denoising, i.e.,
\begin{equation}
\begin{split}
    \pi_\theta(\mathbf{s}, \mathbf{b}) &= p_\theta(\mathbf{b}_{0:T}|\mathbf{s}) \\
    & = \mathcal{N}(\mathbf{b}_T; 0, \boldsymbol{I})\prod_{t=1}^{T}p_\theta(\mathbf{b}_{t-1}|\mathbf{b}_{t}, \mathbf{s}).
\end{split}
\end{equation}
Recall that the definition of $p_\theta(\mathbf{b}_{t-1}|\mathbf{b}_{t}, \mathbf{s})$ has been shown in Eq. (4).
Based on Eq. (4), the probability of each denoising step can be derived as \cite{DDPM}
\begin{equation}
    \mathbf{b}_{t-1} = \frac{1}{\sqrt{\alpha_t}} \left(\mathbf{b}_t - \frac{\beta_t}{\sqrt{1-\bar{\alpha_t}}}\epsilon_{\theta}\left(\mathbf{x}_{t}, t\right)\right) + \sigma_t\mathbf{z},
\end{equation}
where $\mathbf{z} \sim \mathcal{N}(0, \boldsymbol{I})$, and according to \cite{DDPM}, $\sigma_t$ = $\beta_t$ can achieve good performance.
Then, we adopt a Double Deep Q-Network (DDQN) learning architecture \cite{DQN} to organize the ADD training.
Specifically, with the action $\mathbf{b}_0$ generated by policy $\pi_\theta(\mathbf{s}, \mathbf{b})$, ADD applies $\mathbf{b}_0$ in the mobile AIGC environment $\mathbf{s}$ and acquires $R(\mathbf{b}_0|\mathbf{s})$.
The solution evaluation network $Q_v$ can help train parameter $\theta$ in $\epsilon_\theta$ and optimize the policy $\pi_\theta(\mathbf{s}, \mathbf{b})$.
To do so, $Q_v$ calculates the Q-value of $P(\mathbf{b}_0|\mathbf{s})$, and the optimal $\theta$ is the one that can lead to the highest expected Q-value.
In this case, the optimal policy generation network can be obtained by
\begin{equation}
    \underset{\epsilon_{\theta}}{\arg \min }\, \mathcal{L}_{\epsilon}(\theta)=-\mathbb{E}_{\boldsymbol{b}_{0} \sim \epsilon_{\theta}}\left[Q_{v}\left(\mathbf{s}, \mathbf{b}_{0}\right)\right].
\end{equation}
The $Q_v$ should be trained to predict the best Q-values, which is achieved by minimizing the Bellman operator \cite{DQN}.
In the proposed ADD, there are two Q-networks to be trained, namely $Q_{v_1}$ and $Q_{v_2}$, with the corresponding target networks $Q_{v_1}^{*}$ and $Q_{v_2}^{*}$, respectively.
Note that the target networks are used to compute the target for the Q-value updates. 
The weights of $Q_{v_1}^{*}$ and $Q_{v_2}^{*}$ are kept fixed for a number of steps and then periodically updated to match the weights of $Q_{v_1}$ and $Q_{v_2}$, respectively. 
By decoupling the targets from the parameters, the learning process can be stabilized.
Based on Eq. (22), the joint optimization of two Q-networks can be expressed as minimizing the following expectation
\begin{equation}
\mathbb{E}_{\mathbf{b}_{0} \sim \pi_{\theta}^{*}}\left[\left\| \begin{array}{l}
\left(R\left(\mathbf{b}^{0}|\mathbf{s}\right)+\gamma \underset{i=1,2}{\min} Q_{v_{i}^{\prime}}\left(\mathbf{s}, \mathbf{b}_{0}\right)\right) \\
-Q_{v_{j}}\left(\mathbf{s}, \mathbf{b}_0\right)
\end{array} \right\|^{2}\right],
\end{equation}
where $\gamma$ is the discount factor; $j \in \{1, 2\}$ equals the value of $i$ that leads to the minimum $Q_{v_{i}^{\prime}}\left(\mathbf{s}, \mathbf{b}_{0}\right)$.
The policy can be optimized with the optimal Q-networks, and the optimal bandwidth allocation scheme $\mathbf{b}$ in any given AIGC state $\mathbf{s}$ can be generated.
The detailed training and inference procedures of ADD are shown in \textbf{Algorithm 2}.

\subsection{Complexity Analysis}
Here, we analyze the complexity of the proposed ADD algorithm.
Suppose that the sizes of the diffusion-based policy network and Q-network are $S_p$ and $S_q$, respectively.
The architectural complexity is $\mathcal{O}(S_p + 2 S_q)$.
Since generating each bandwidth allocation scheme requests $T$ times diffusion denoising, the policy generation complexity is $\mathcal{O}(TS_p)$.
Hence, the overall complexity can be derived as $\mathcal{O}((T + 1)S_p + 2S_q)$.
Accordingly, supposing that $\delta$ epochs are performed, and the batch size is $S_b$, the computational complexity for training is $\mathcal{O}(\delta S_b((T + 1)S_p + 2S_q))$.
Finally, the corresponding inference-stage complexity is $\mathcal{O}(S_p)$.

\begin{figure*}[tbp]
\centerline{\includegraphics[width=1.98\columnwidth]{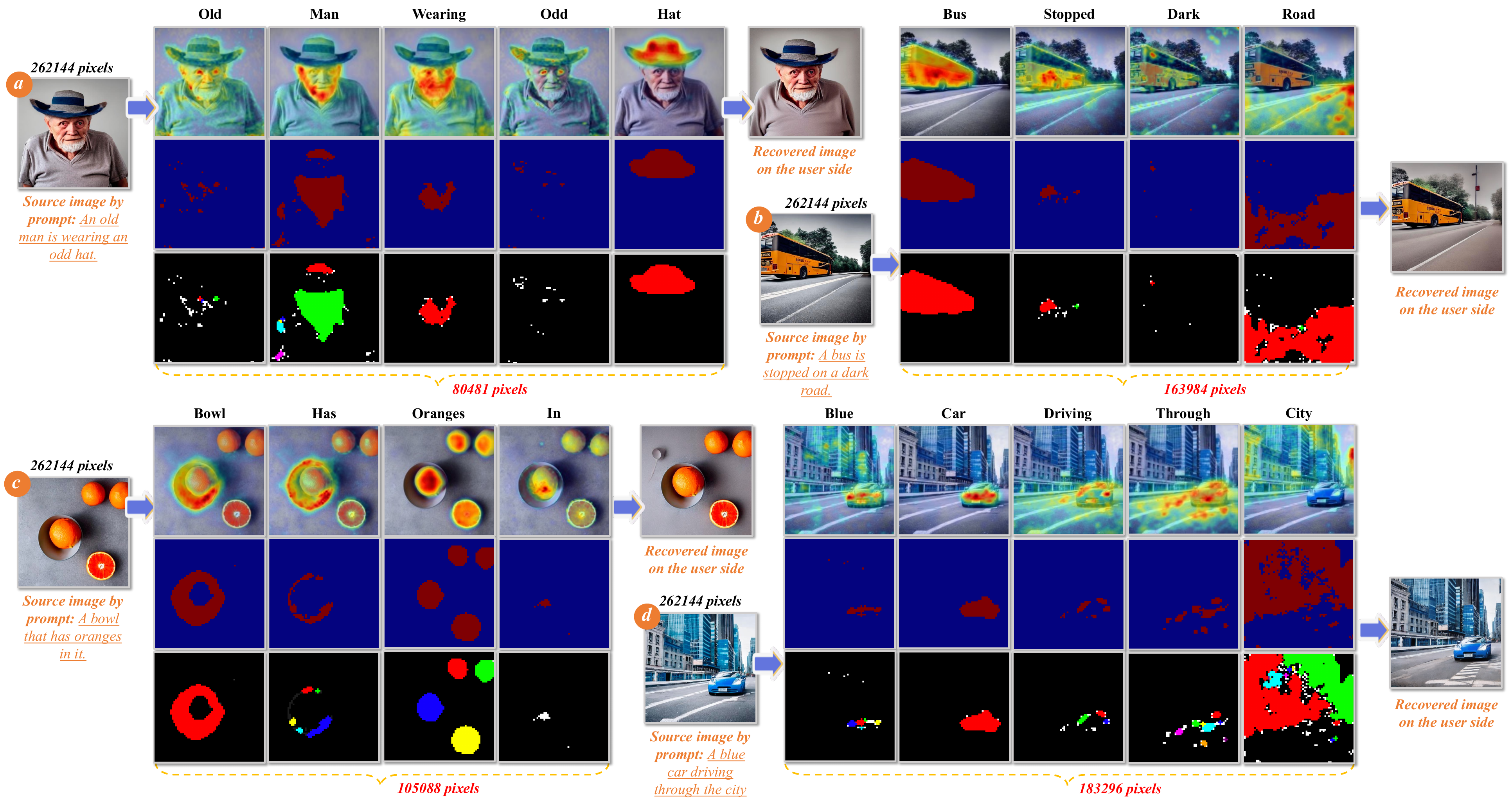}}
\caption{The case study to illustrate the effectiveness of G-SemCom-aided mobile AIGC. \texttt{(a)}: The case of prompt: \textit{An old man is wearing an odd hat}. \texttt{(b)}: The case of prompt: \textit{A bus is stopped on a dark road}. \texttt{(c)}: The case of prompt: \textit{A bowl that has orange in it}. \texttt{(d)}: The case of prompt: \textit{A blue car driving through the city}. For each case, the three rows are cross-modal attention maps, binary attention maps, and clustering results, respectively.}
\label{exp1}
\end{figure*}
\section{Performance Evaluation}
In this section, we implement the proposed G-SemCom framework and build the experimental mobile AIGC system.
Then, we conduct extensive experiments that aim to answer two questions: 1) whether the proposed G-SemCom framework for mobile AIGC can effectively reduce the bandwidth consumption of users while ensuring them acquire high-quality AI-generated images and 2) whether the ADD algorithm can efficiently allocate bandwidth resources among users, thus maximizing the overall reward defined by JPSQ.
The analysis of the experimental results is also described.

\textbf{Implementation}. 
To generate high-quality source images, we equip MASPs with Stable Diffusion v2 \cite{SDpaper}, the state-of-the-art text-to-image AIGC model.
The number of diffusion steps is set to 25.
On the user side, we deploy a diffusion-based image inpainting model \cite{Inpainting1} as the generative decoders.
Since the users only need to recover the image background with less semantic importance, the diffusion step number is set to 5 to reduce resource consumption.
The prompts that the users send to the MASP are selected from the image captions in the \textit{COCO 2017} dataset \cite{COCO}.
We adopt the implementation of the DreamSim metric in \cite{DreamSim}, which ensembles CLIP, OpenCLIP, and DINO to construct the backbone model. 
For NIMA, we utilize MobileNet to implement the image quality classifier and load the pretrained model weights from \cite{NIMAmobile}.
All the steps of \textbf{Algorithm 1} are packed into a pipeline written in Python, based on \textit{diffusers}, \textit{daam}, \textit{spacy}, and \textit{sklearn} libraries. 
Finally, we leverage \textit{PyTorch} to implement the proposed ADD algorithm, combining the basic DDQN architecture in \cite{DDQN2} and our deep diffusion module for policy generation.

\textbf{Testbed}. The experiments are conducted on a server with an NVIDIA RTX A5000 GPU with 24 GB of memory and an AMD Ryzen Threadripper PRO 3975WX 32-Core CPU with 263 GB of RAM. The operating system is Ubuntu 20.04 LTS with \textit{PyTorch} 2.0.1. We utilize this server to simulate one MASP and multiple uniformed distributed mobile users. The OFDMA transmission model between the MASP and users is built based on \cite{Yining}.
\renewcommand{\arraystretch}{1.2}
\begin{table}
\begin{tabular}{l|p{4.8cm}|p{1.8cm}<{\centering}}
\Xhline{2.2pt}
\rowcolor[rgb]{0.92,0.92,0.92}
\textbf{Symbol}&\multicolumn{1}{c|}{\textbf{Description}}&\textbf{Value}\\
\hline
$\gamma$ & Discount factor of ADD& 0.95\\
\hline
$L_i$& Latency threshold of $U_i$& 5s\\
\hline
$O$ & Bandwidth consumption for transmitting each pixel& 1 \\
\hline
$Q_\mathrm{th}$ & Threshold of aesthetic quality& 4.9827\\
\hline
$\eta$ & Learning rate of ADD& $10^{-4}$\\
\hline
$T$ & Diffusion step number of ADD& 5 \& 6\\
\hline
$\xi$& Threshold of attention value& {0.9, 0.8, 0.5}\\
\hline
$\omega_0$& Weighting factor in Eq. (17)& 1.25 \\
\hline
$\omega_1$& Weighting factor in Eq. (19)& 500 \\
\hline
$\omega_2$& Weighting factor in Eq. (19)& 0.05 \\
\Xhline{2.2pt}
\end{tabular}
\caption{The summary of experimental settings \cite{Yining}.}
\end{table}
\renewcommand{\arraystretch}{1}

\textbf{Experimental Settings}. 
The important environmental and hyperparameter configurations in terms of the proposed G-SemCom pipeline and ADD are shown in TABLE III.

\subsection{Effectiveness of G-SemCom Framework}
In this part, we evaluate the effectiveness of the proposed G-SemCom framework for mobile AIGC.

\subsubsection{Case Study}
First, Fig. \ref{exp1} illustrates four cases where the users request different images from the MASP.
We can observe that the cross-modal attention maps can effectively associate any given word to certain pixels of the generated images.
However, the attention distribution of $\textit{\textbf{ADJ}}$- and $\textit{\textbf{VERB}}$-typed words, e.g., \texttt{[old]} in case \textit{a} and \texttt{[stopped]} in case \textit{b}, are scattered due to less clear semantic meaning.
Take \texttt{[old]} as an example.
Its attention covers the entire image, while only those pixels highlighting the character's facial features are meaningful.
To this end, we utilize Eq. (14) to filter out attention points without clear semantic meaning, forming the binary attention maps.
Note that we assign $\xi$ in Eq. (14) with different values according to the general semantic importance of different part-of-speech types.
Specifically, for $\textit{\textbf{PROPN}}$-typed, $\textit{\textbf{NN}}$-typed, and other words, $\xi$ equals \textit{0.9}, \textit{0.8}, and \textit{0.5}, respectively.
We can observe that the meaningless pixels are effectively removed.
In contrast, the pixels with strong semantic meaning, such as the pixels associated with words \texttt{[hat]} and \texttt{[bus]}, are fully maintained.
Atop binary attention maps, the proposed attention clustering algorithm based on DBSCAN can further remove the noise and noisy clusters, which are too small and will affect image recovery.
Finally, the recovered images can hold the high quality of source images, with almost no perceptual quality difference.
Meanwhile, the semantic information takes only 80481, 163984, 105088, and 183296 pixels compared with the 512\,$\times$\,512 source image with 262144 pixels, achieving 69.3\%, 27.4\%, 60.0\%, and 30.1\% reduction, respectively.
More in-depth experiments on bandwidth reduction are shown below.
\begin{figure}[tbp]
\centerline{\includegraphics[width=0.98\columnwidth]{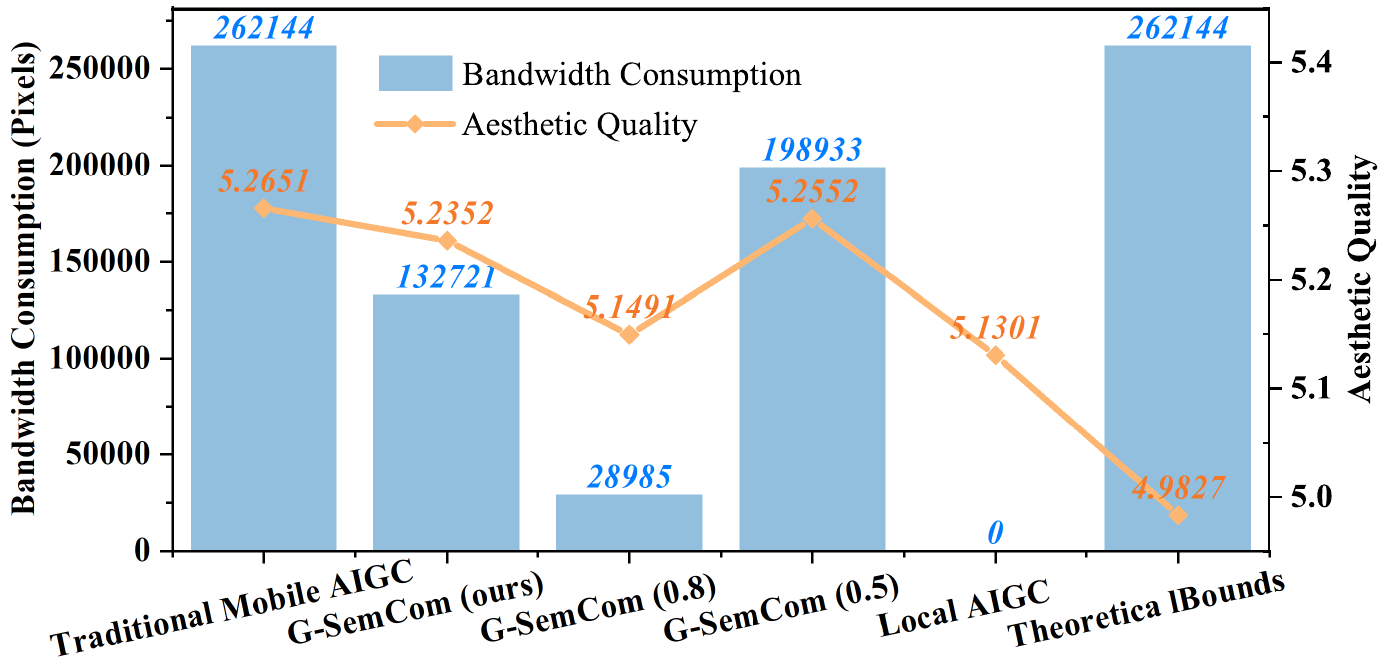}}
\caption{The performance of different AIGC paradigms in terms of the bandwidth consumption and image quality. Note that \textit{Bounds} include the theoretical upper and lower bounds of the bandwidth consumption and NIMA score, respectively.}
\label{G-performance}
\end{figure}
\begin{figure}[tbp]
\centerline{\includegraphics[width=0.85\columnwidth]{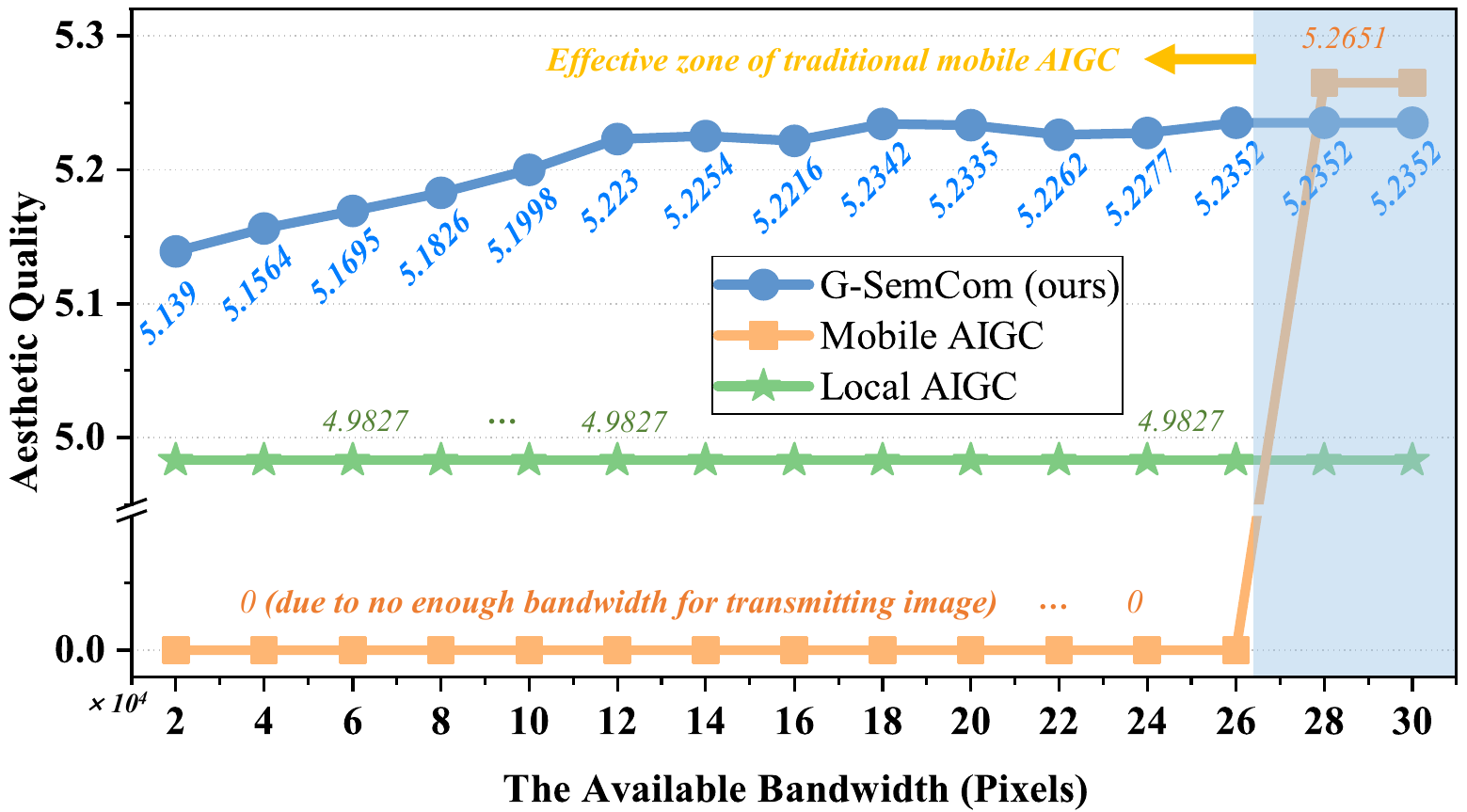}}
\caption{The service robustness of difference AIGC paradigms to channel error. Note that the blue bar indicates the effective zone of mobile AIGC since it can only provide the users with useful images within this range.}
\label{robust}
\end{figure}

\subsubsection{G-SemCom Performance}
Fig. \ref{G-performance} illustrates the average performance of the traditional and our proposed G-SemCom-aided mobile AIGC for generating 1000 images.
Note that we take the number of pixels as the bandwidth unit, which can circumvent the errors caused by different standards for packing images, e.g., \textit{jpg} and \textit{png} \cite{JPG}.
From Fig. \ref{G-performance}, we can observe that compared with traditional mobile AIGC, the proposed G-SemCom [with $\xi$ = \{0.9, 0.8, 0.5\}] can reduce the bandwidth consumption of the users by 49.4\% on average, while the average image quality, measured by NIMA score, drops only 0.0299.
Moreover, if $\xi$ is relaxed to 0.8 or tightened to 0.5, the bandwidth and quality will further decrease and increase, respectively, exhibiting the outstanding flexibility of our proposal.
Recall that the generative decoder on the user side is a lightweight image inpainting model, which inpaints the masked image with 5 diffusion steps.
We then provide fully masked images to the decoder, thereby exploring the quality of the images that the users can generate locally using the same computation resources with G-SemCom.
As shown in Fig. \ref{G-performance}, the image quality of local AIGC is only 5.1301 due to less powerful models and fewer diffusion steps.
Given the lower bound of the NIMA score is 4.9827, the decrement from 5.2651 to 5.1301 means that the image quality drops by 47.8\%. 
Note that such a lower bound is acquired by generating 1000 pure Gaussian noise images, meaning their NIMA scores, and taking the minimum value.

\subsubsection{Service Robustness}
Besides saving bandwidth, another significant advantage of our G-SemCom framework is enhancing the robustness of mobile AIGC services to channel errors.
Note that channel error means that the connection between the user and MASP is interrupted due to unexpected circumstances, resulting in only a part of the semantic information being transmitted.
Traditionally, users need to download the entire image from the MASP.
Hence, the channel error will cause transmission failure, in which the users can only receive broken images.
Assisted by G-SemCom, regardless of how many bits have been transmitted, users can recover full images using the generative decoders, which is extremely important if the application has strict requirements for latency and cannot tolerate re-transmission. 
As shown in Fig. \ref{robust}, even though only 20000 pixels are transmitted, the average NIMA score reaches 5.1390, which exceeds that of the local AIGC (i.e., 5.1301).
With the increasing number of transmitted pixels, the quality of the recovered images grows gradually.
In contrast, the image quality of traditional mobile AIGC remains 0 until the entire image information can be transmitted.
Apart from enhancing service robustness, the proposed ADD can further determine the number of pixels to be transmitted for achieving the best JPSQ-bandwidth balance in the given state.
The corresponding experiments are discussed in Section VI.B.
\begin{figure}[tbp]
\centerline{\includegraphics[width=0.85\columnwidth]{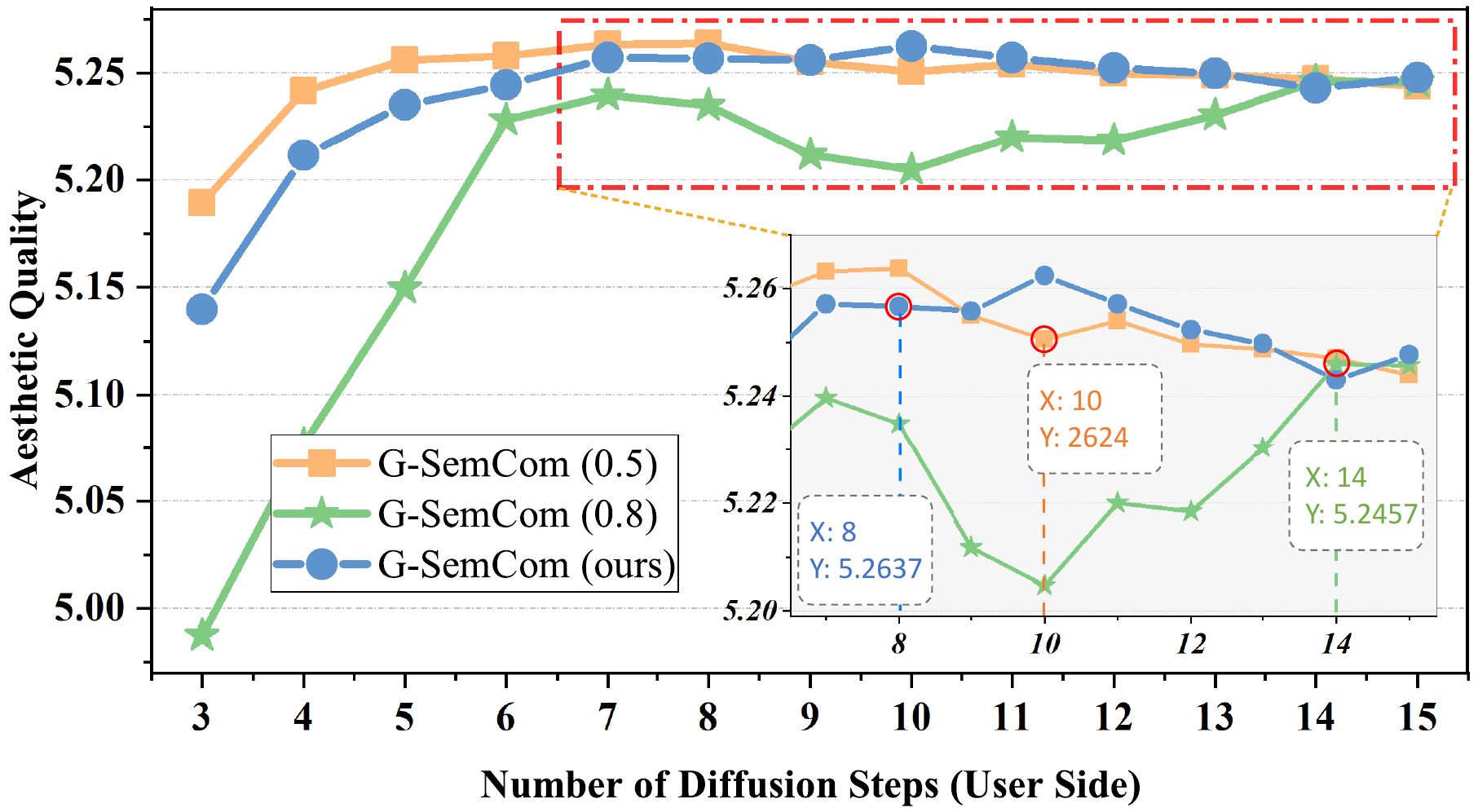}}
\caption{The aesthetic quality with the number of diffusion steps on the user side.}
\label{step}
\end{figure}
\renewcommand{\arraystretch}{1.2}
\begin{table}[tpb]
\centering
\begin{tabular}{l|p{1.8cm}<{\centering}|p{1.8cm}<{\centering}|p{1.8cm}<{\centering}}
\Xhline{2.2pt}
\rowcolor[rgb]{0.92,0.92,0.92}
\textbf{Stakeholder}&\multicolumn{1}{c|}{\textbf{GPU time}}&\textbf{CPU time}&\textbf{GPU memory}\\
\hline
MASP& 2.955s& 3.181s & 7846 MB \\
\hline
User& 1.078s& 1.170s & 3919 MB  \\
\Xhline{2.2pt}
\end{tabular}
\caption{The computation and memory consumption of the MASP and users.}
\end{table}
\renewcommand{\arraystretch}{1}
\begin{figure}[htpb]
\centerline{\includegraphics[width=0.96\columnwidth]{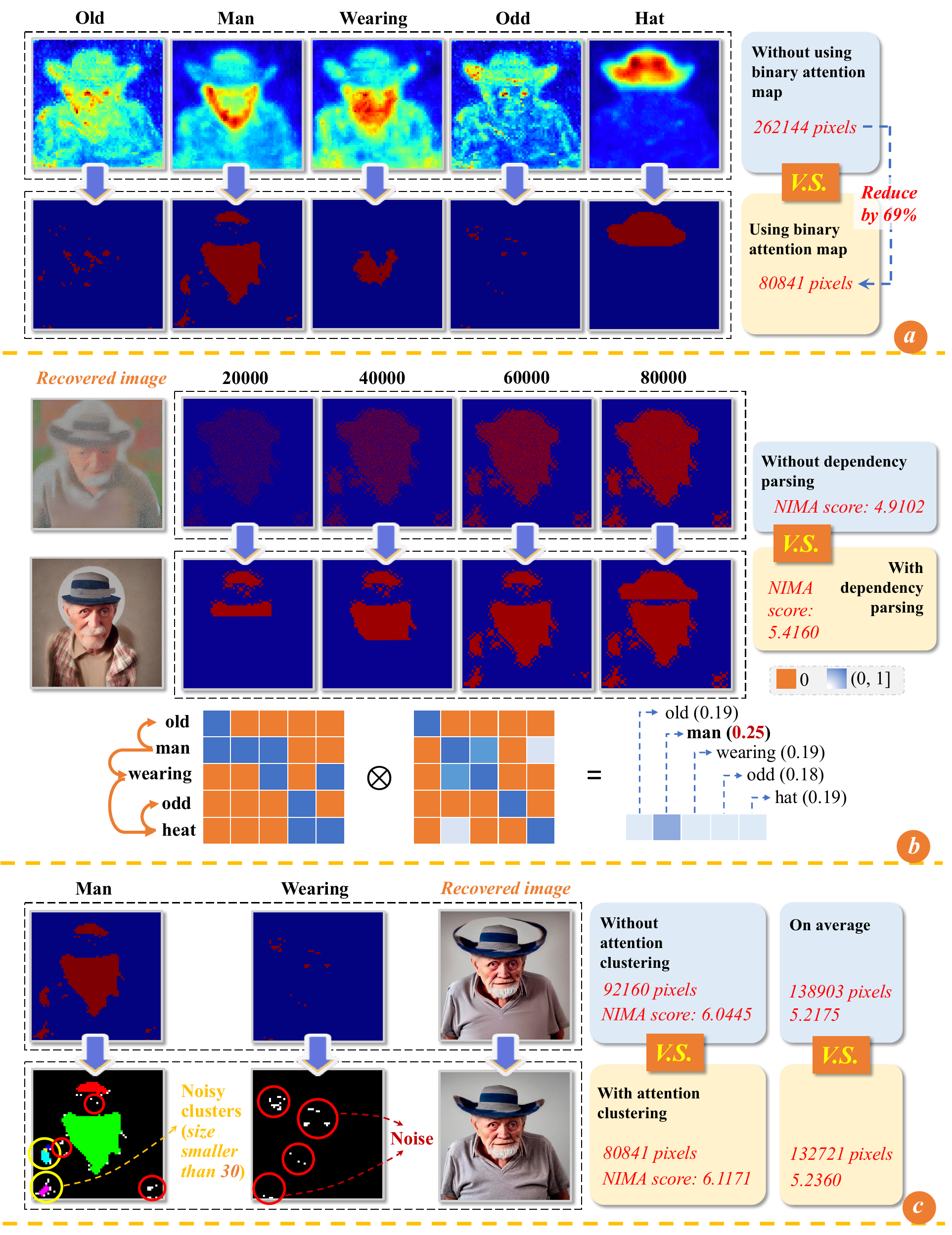}}
\caption{The ablation study. \texttt{(a)}: The inspection on binary attention map. \texttt{(b)}: The inspection on dependency parsing. \texttt{(c)}: The inspection on attention clustering.}
\label{abtest}
\end{figure}
\begin{figure}[htbp]
\centerline{\includegraphics[width=0.92\columnwidth]{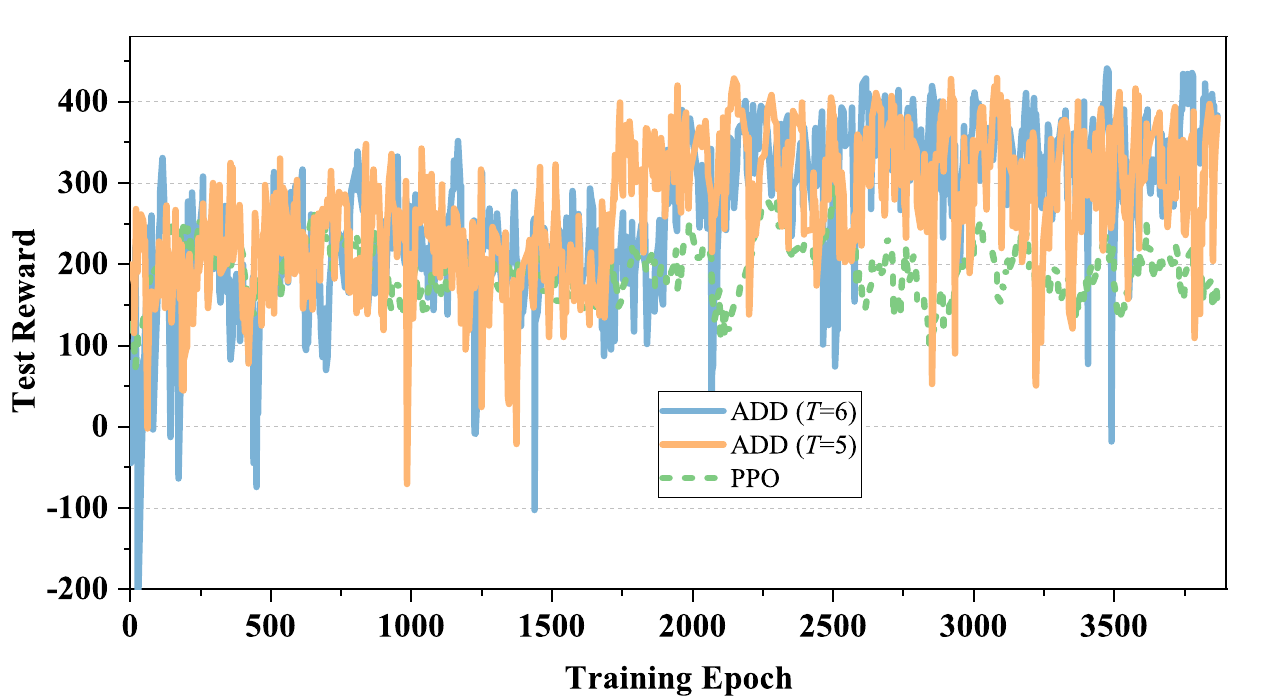}}
\caption{The training curves of the proposed ADD (with $T$ = 5 and 6) and baseline PPO.}
\label{training}
\end{figure}

\subsubsection{Resource Consumption on the User Side}
Up till now, we have demonstrated that the proposed G-SemCom framework can help users greatly save bandwidth resources while receiving high-quality images and robust mobile AIGC services.
Here, we explore the resource consumption of users to operate the generative decoder.
To this end, we first evaluate the relationship between diffusion steps adopted by generative decoders and the image quality.
As shown in Fig. \ref{step}, the increasing diffusion step number fails to improve the image quality linearly because the pixels yet to be recovered only have weak semantics.
Take case \textit{a} in Fig. \ref{exp1} as an example. 
Compared with the man's face that accommodates rich semantics (e.g., facial features, expression, and beard), human perception is much less sensitive to the cloth and background.
Therefore, increasing resources for rendering such parts cannot improve the human-perceptual quality of the recovered image.
For the recommended $\xi$ scheme, i.e., \{0.9, 0.8, 0.5\}, setting the diffusion step number as 5, 6, or 7 can already lead to satisfying image quality.
TABLE IV illustrates the resource overhead when the diffusion step number equals 5.
We can observe that compared with the MASP, which needs to perform the complete AIGC inferences locally, the users only need to consume 63.4\%, 63.3\%, and 51.1\% of CPU time, GPU time, and GPU memory, respectively.
The current user-side mobile devices, e.g., smartphones and tablets, can well afford such levels of resource consumption \cite{chen2023speed}.
Moreover, the MASP can keep evolving its AIGC model to improve generation quality while users can use the same lightweight decoder for recovery.
Such a paradigm ensures the outstanding sustainability of the G-SemCom-aided mobile AIGC to fit the fast advancements of AIGC models in the future.

\subsection{Ablation Study}
In this part, we perform an in-depth ablation study, aiming to investigate the effectiveness of each proposed step performed by the MASP in \textbf{Algorithm 1}.

\subsubsection{Binary Attention Map}
This operation refers to filtering the less important attention points from the original attention maps, whose major purpose is reducing semantic information size. Fig. \ref{abtest}(a) illustrates the case \textit{a} in Fig. \ref{exp1} with and without filtering. Note that we adopt the recommended $\xi$ scheme, i.e., \{0.9, 0.8, 0.5\} when constructing the binary attention maps. We can observe that without filtering, the attention maps contain all the pixels, while the majority of them only have weak semantic meaning. In contrast, by setting the threshold $\xi$ and forming the binary attention map, the size of the semantic information can be reduced by 69.3\%.

\subsubsection{Dependency Parsing}
Dependency parsing facilitates the MASP to evaluate the importance of each word in the user prompts. Without dependency parsing, the MASP can only randomly select pixels when packing the semantic information. As shown in Fig. \ref{abtest}(b), the resulting semantic information is blurred, with a NIMA score of 4.9102. In contrast, we first understand which pairs of words are correlated by dependency parsing. For instance, Fig. \ref{abtest}(b) shows the dependencies existing in our case. Then, the $\boldsymbol{C}^{*}$ and $\boldsymbol{D}^{*}$ matrices can be established. Finally, the semantic importance can be calculated, which is $\mathbf{s}$ = [0.19, 0.25, 0.19, 0.18, 0.19]. The packing of semantic information can then follow the order of $\mathbf{s}$. We can observe that the pixels associated with \texttt{[man]} are prioritized. In this case, even though only 20000 pixels are transmitted, the core semantic information is well preserved in the recovered image, resulting in a much higher NIMA score.
\begin{figure}[tbp]
\centerline{\includegraphics[width=0.95\columnwidth]{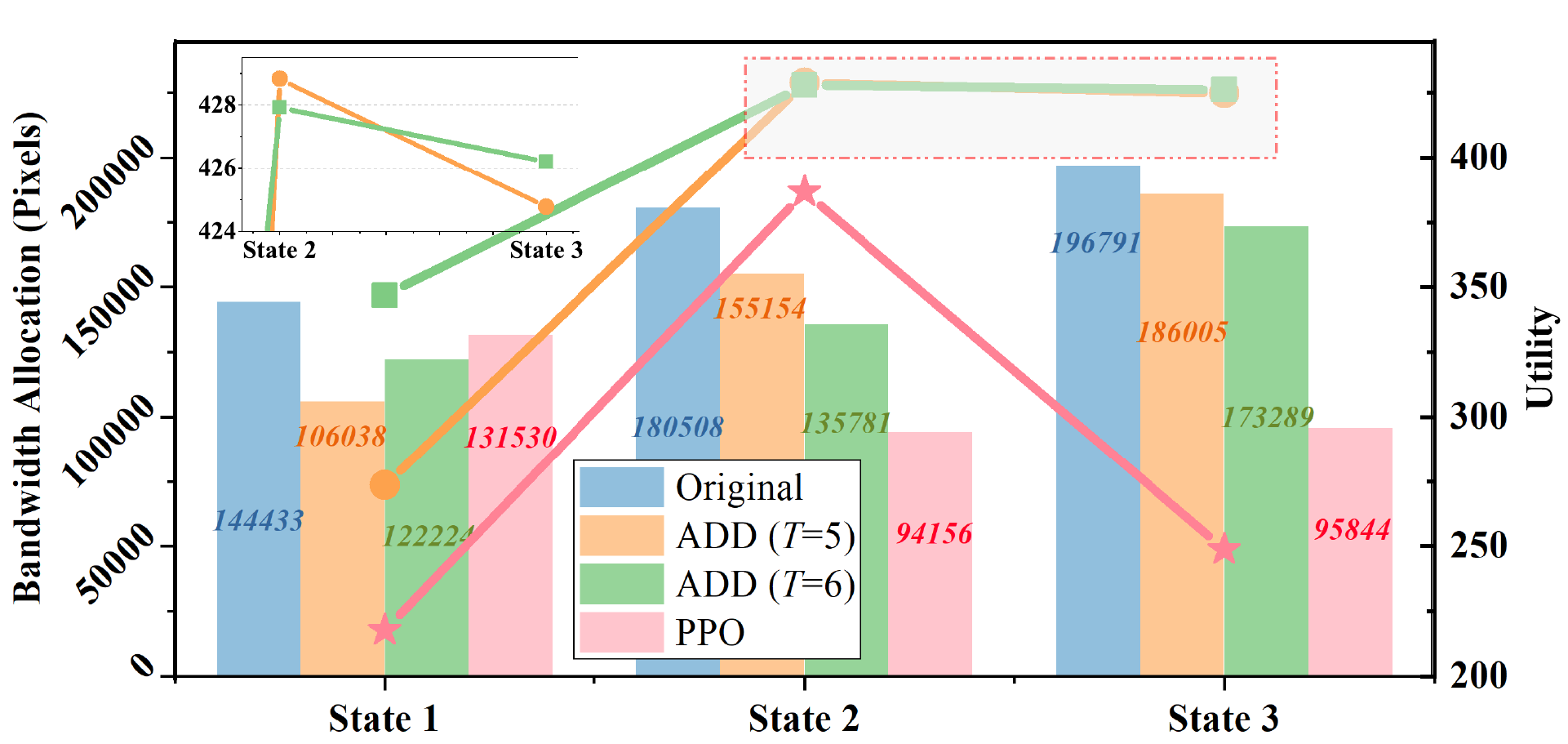}}
\caption{The bandwidth allocation schemes and resulting reward in different states. Note that \textit{state} represents the semantic information $\boldsymbol{S}$ of a mobile user.}
\label{one}
\vspace{0.7cm}
\centerline{\includegraphics[width=0.92\columnwidth]{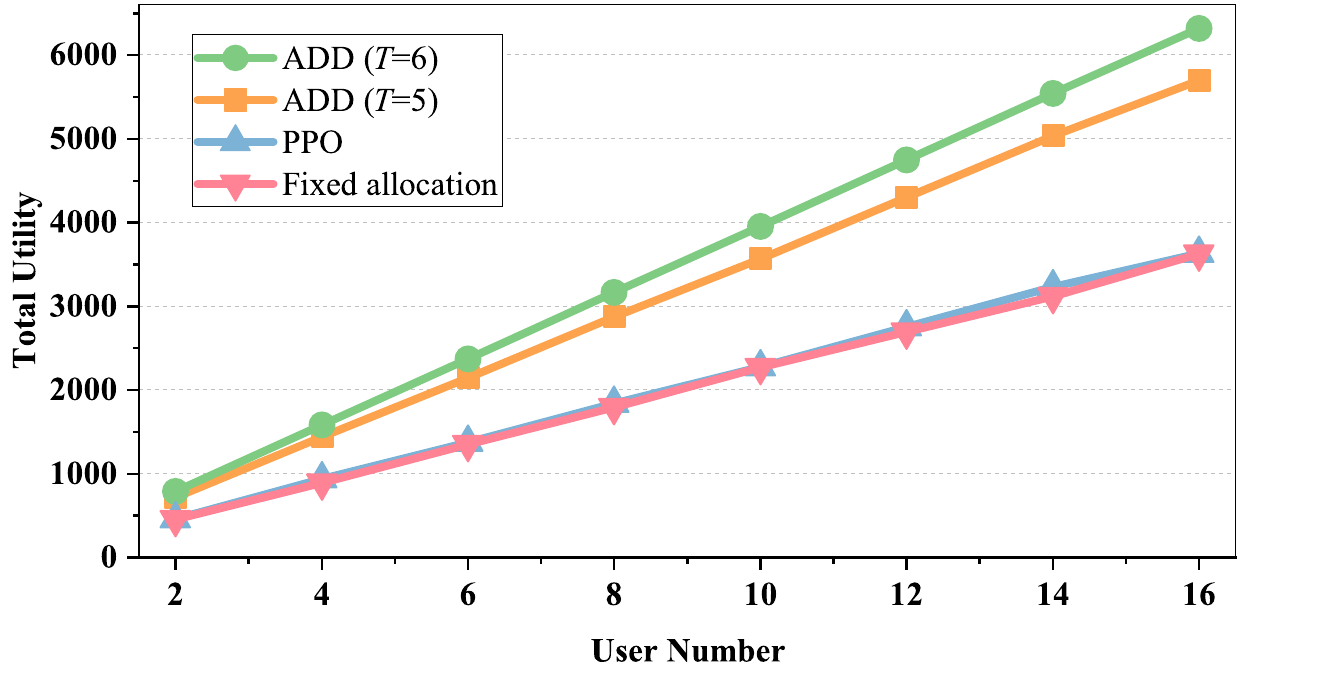}}
\caption{The total utility with increasing user number. \textit{Fixed allocation} means allocating bandwidth for the entire semantic information without optimization.}
\label{total}
\end{figure}

\subsubsection{Attention Clustering}
Finally, attention clustering means adopting DBSCAN to remove noise and noisy clusters, which do not carry enough image semantics due to small sizes. 
Moreover, they might affect image recovery since the generative decoder can hardly generate. 
Fig. \ref{abtest}(c) illustrates the clustering results of \texttt{[man]} and \texttt{[wearing]}. The pixels marked by red and yellow circles are noise and noisy clusters, respectively. We can observe the noise associated with the word \texttt{[wearing]} affects the recovery of the man's hat, decreasing the image quality from 6.1171 to 6.0445. Using the 1000 images of Fig. \ref{exp1}, by performing attention clustering, the average bandwidth consumption reduces by 6182 pixels while the quality increases by 0.0185.

\subsection{Efficiency of ADD}
Then, we study the efficiency of the proposed ADD algorithm in optimizing the resource allocation scheme.

\subsubsection{ADD Training}
First, Fig. \ref{training} illustrates the training curves of the ADD algorithm with 5 and 6 steps of diffusion.
Similar to Section VI-A, the training dataset is constructed by images generated from captions in \textit{COCO 2017} dataset \cite{COCO}.
Note that the JPSQ calculation is time-consuming since two AI inferences are performed for acquiring DreamSim and NIMA, respectively.
To this end, we constrict an action-reward table containing the JPSQ score of taking each action on each training image to facilitate the training.
Moreover, to demonstrate the superiority of our proposal, we adopt a standard DRL algorithm as the baseline, called PPO \cite{PPO_add}.
As shown in Fig. \ref{training}, the ADD ($T$=5) and PPO take a similar time to converge.
The ADD ($T$=6) converges slower while achieving almost 90\% higher rewards than PPO.
This can be explained by the enhanced ability of ADD to explore the environment since an exploration noise is added to the generated bandwidth allocation scheme during each training iteration.
Hence, the training process can avoid getting trapped in sub-optimal solutions.
However, the number of diffusion steps is not always better. 
In our case, the training will be difficult to converge if generating each bandwidth scheme performs 7 or more times of diffusion denoising.
This is because ADD will lose its ability to explore the state effectively, as excessive denoising might lead to overfitting. 

\subsubsection{Optimization of JPSQ}
Then, we investigate the efficiency of ADD in scheduling bandwidth.
Fig. \ref{one} shows the bandwidth allocation schemes for three images whose semantic information sizes are 144433, 180508, and 196791 pixels, respectively.
We can observe that the proposed ADD algorithm with $T$ = 5 and 6 can achieve 32.2\% and 40.9\% higher utility [defined in Eq. (19)] than PPO on average.
Such results demonstrate that our algorithm can better balance the human-perceptual AIGC service quality and bandwidth costs.
Take State \textit{3} as an example. 
The PPO algorithm only assigns 94156 pixels for transmitting the semantic information.
In this case, even though bandwidth consumption is low, the DreamSim score is 0.198, meaning the recovered image holds 80.2\% similarity with the source image at the semantic level.
In contrast, the ADD ($T$ = 6) algorithm uses 173289 pixels to achieve 96.7\% similarity and a 5.56 NIMA score, resulting in 72.5\% higher overall utility.
Furthermore, we evaluate the total utility with the increasing number of users, as shown in Fig. \ref{total}.
From Fig. \ref{total}, we can observe that the ADD algorithm with $T$ = 5 and 6 significantly outperforms the PPO and default schemes. 
Efficient bandwidth allocation is crucial for mobile AIGC to meet users' demand for high-quality AIGC outputs with limited bandwidth resources.

\section{Conclusion}
In this paper, we have presented a novel G-SemCom framework for mobile AIGC, where the MASP only sends compressed semantic information, and users adopt a lightweight generative decoder to recover high-quality images.
Specifically, we have introduced cross-modal attention maps to map the semantics between different modalities.
Assisted by attention maps, the MASP can perform dependency parsing on user prompts and find the pixels with the strongest semantic importance as the semantic information.
Moreover, considering the bandwidth limitation of mobile environments, we have defined a joint optimization problem to optimize the bandwidth allocation among users.
Particularly, noting that semantic information not only preserves the image semantics but also serves as the prompt for image recovery, we have presented a human perceptual metric named JPSQ, combining two learnable measurements regarding image semantic similarity and aesthetic quality, respectively.
Furthermore, leveraging attention maps and the diffusion principle, we have efficiently designed the ADD algorithm to maximize the JPSQ.
Extensive experiments demonstrate that our G-SemCom framework can reduce bandwidth consumption by 49.4\% while ensuring image quality on the user side.
In addition, the ADD has significantly outperformed traditional DRL, striking great balances between bandwidth and JPSQ in mobile environments.

Looking to the future, the G-SemCom framework can further assist mobile AIGC via fine-grained semantic analysis of the AIGC outputs empowered by cross-modal attention maps.
For instance, the MASP can clean the AIGC outputs to safeguard users, especially children, from inappropriate or harmful content created by AIGC models.
Moreover, with the ever-increasing complexity of mobile AIGC tasks, the one-round service paradigm can hardly meet user requirements. 
Our framework can enable the interactive AIGC services. 
Specifically, users can evaluate the received AIGC product, mark the prompt words that are not illustrated properly, and prompt the refinements. 
Then, the MASP can easily locate the corresponding parts via attention maps and fine-tune them by controlled generation.
Such directions are critical for mobile AIGC and are worth further research.


\ifCLASSOPTIONcaptionsoff
  \newpage
\fi



%
\bibliographystyle{IEEEtran}
\bibliography{gsemcom}
\vfill

%




\end{document}